\documentstyle[epsfig]{elsart}
\newcommand{\be}{\begin{equation}}
\newcommand{\ee}{\end{equation}}

\def\simlt{\lower.5ex\hbox{\ltsima}}
\def\gtsima{$\; \buildrel > \over \sim \;$}
\def\simgt{\lower.5ex\hbox{\gtsima}}

\def\simlt{\lower.5ex\hbox{\ltsima}}
\def\gtsima{$\; \buildrel > \over \sim \;$}
\def\simgt{\lower.5ex\hbox{\gtsima}}
\def\hmpc{{\rm\,h^{-1} Mpc}}

\def\msunh{{\rm\,h^{-1}~M_\odot}}

\def\cm{{\rm\,cm}}

\def\ergcm2{\ {\rm erg~cm^{-2} }}
\def\ergscm2{\ {\rm erg~s^{-1}~cm^{-2} }}
\def\hmpc{\ h^{-1}~{\rm Mpc} }
\def\cm2s{\ cm^2 ~s^{-1} }

\def\s{\ifmmode \widetilde \else \~\fi}
\def\={\overline}

\def\spose#1{\hbox to 0pt{#1\hss}}

\def\eg{{\it e.g.\ }}

\def\ie{{\it i.e.\ }}

\def\lta{\mathrel{\spose{\lower 3pt\hbox{$\mathchar"218$}}
     \raise 2.0pt\hbox{$\mathchar"13C$}}}
\def\gta{\mathrel{\spose{\lower 3pt\hbox{$\mathchar"218$}}
     \raise 2.0pt\hbox{$\mathchar"13E$}}}
\def\mincir{\ \raise -2.truept\hbox{\rlap{\hbox{$\sim$}}\raise5.truept  
\hbox{$<$}\ }}                                                          %
\def\magcir{\ \raise -2.truept\hbox{\rlap{\hbox{$\sim$}}\raise5.truept  %
\hbox{$>$}\ }}                                                          %
\def\simlt{\ \raise -2.truept\hbox{\rlap{\hbox{$\sim$}}\raise5.truept   
\hbox{$<$}\ }}                                                          %
\def\simgt{\ \raise -2.truept\hbox{\rlap{\hbox{$\sim$}}\raise5.truept   %
\hbox{$>$}\ }}                                                          %
\def\newline{\par\noindent}

\def\ea{{\it et al.} \,}

\def\s-z{S-Z}

\begin{document}
\begin{frontmatter}
\title{CLUSTERS OF GALAXIES\\
       AND THE DIFFUSE GAMMA RAY BACKGROUND}

\author[Roma]{Sergio Colafrancesco} and
\author[Chicago]{Pasquale Blasi}
\address[Roma]{Osservatorio Astronomico di Roma,
Via dell'Osservatorio 2, I-00040 Monteporzio, Italy}
\address[Chicago]{Dept. of Astronomy \& Astrophysics, E. Fermi Inst.,
University of Chicago, Chicago, IL 60637-1433, USA}

\begin {abstract}
We discuss the diffuse emission of gamma rays and neutrinos from 
galaxy clusters in the viable models for structure formation in the
universe.
We use a self-consistent picture for cluster formation and evolution
starting from a primordial density perturbation spectrum, and a
realistic modelling for the distribution of the intergalactic medium
which is abundantly present within galaxy clusters.
We find that an evolving population of clusters can produce a fraction
$\sim 0.5 \div 2 \%$ of the diffuse gamma-ray background
(DGRB) observed by EGRET.
This result is robust and is weakly dependent on the cosmological
scenario and on the degree of
evolution of the inter galactic medium (IGM)
in distant clusters, because the bulk of the 
sources contributing to the DGRB is located at redshifts 
$z \simlt 0.2$.
We also found a correlation between the non-thermal,
gamma-ray and the thermal X-ray emissions from these structures.
Using this result, we 
derived a list of gamma-ray clusters observable with the
next generation $\gamma$-ray detectors. 
Finally, we briefly discuss the possible
relevance of galaxy clusters for neutrino
astronomy and for very high energy particle astronomy. 
\end{abstract}
\end{frontmatter}

\section{Introduction}
The EGRET experiment on board CGRO
 revealed the existence of a diffuse gamma-ray 
background (hereafter DGRB) at the level of 
$I_{DGRB} = 9.6 \cdot 10^{-7}
E_{GeV}^{-2.11 \pm 0.05} ~ cm^{-2} s^{-1} sr^{-1} GeV^{-1}$ 
\cite{owz94}
in the energy range $0.03 \div 10$ GeV.
However, a recent reanalysis of the EGRET data 
\cite{swz}
found that the level of the DGRB is 
systematically lower by a factor $\sim 20 \%$ in the energy range
$\sim 0.1 \div 4$ GeV.
The DGRB is observed at high galactic latitudes $b > 10$ deg
and such an evidence suggested an extragalactic origin for this 
diffuse background.
Nonetheless, the specific origin of the DGRB is still under debate.
In fact, the EGRET experiment \cite{kan}
has a poor angular resolution ($\theta_{min} 
\sim 1$ deg) 
so that it is hard to discriminate among different origins of this
extragalactic background.
Specifically, it is still difficult to discriminate between a purely 
diffuse nature of
the DGRB (see \eg \cite{ca})
and the option of a DGRB  made by a superposition of unresolved, 
discrete sources.

The large number of identified AGNs and flat spectrum radio quasars 
(hereafter FSRQ)
in the EGRET sky (\cite{fi96} \cite{has} \cite{mat}) 
suggested
that most of the DGRB can be produced by a non-resolved population
of AGNs, the actual fraction of the DGRB produced by FSRQ and BLLacs 
being in the range $\sim 40  \div  95 \%$ 
(see \eg \cite{com}).
Separately, it has been evaluated that a fraction $\sim 42 \div 97 \%$ 
of the DGRB could be ascribed to blazars
\cite{pad}.
However, the flatness of
the spectrum of the DGRB seems to favour 
the possibility that BLLacs could be
the major contributors to the DGRB of extragalactic origin \cite{pohl}.
Erlykin \ea \cite{erl96} reviewed the various AGN contributions and
quoted that the fraction of the DGRB produced by the observed AGNs is
$\sim 65 \%$.
The DGRB fractions previously reported may be subject to a revision
($\sim 25 \%$ increase) if the recent reanalysis \cite{swz} of the
EGRET data is adopted.

On account of the large theoretical uncertainties 
and of the present observational precision of the EGRET detectors, 
it is still hard to discriminate
among the different proposed possibilities, even though a fluctuation
analysis of the EGRET data should give more precise indications on the
nature and origin of the DGRB.

Beside the discrete, unresolved source case pictured for the origin of
the DGRB,
there have been some pioneering works \cite{hw}
\cite{ds} \cite{bbp} \cite{volk}
suggesting that a relevant
fraction of the DGRB could be produced by {\it extended} sources 
through
hadronic collisions of cosmic ray (hereafter CR) protons
interacting with the protons of the Inter
Galactic Medium (hereafter IGM) which is abundantly present within 
galaxy clusters (see \cite{sa88}
for a review).
In this alternative picture, the CR's are assumed to be produced within 
clusters (we will discuss in Sect.4 some of the possible sources)
where also a population of protons and electrons is residing in the form of
a hot (with temperatures $T \sim 10^7 \div 10^8$ K), tenuous (with
electron number densities $n_e \sim
10^{-3}~cm^{-3}$), chemically enriched and massive 
(with mass fractions $M_{IGM}/M \sim 0.05 \div 0.3$) plasma: the IGM.

The proposed mechanism has an essential ingredient in the confinement 
of the CR's within clusters where they are produced; this point, already 
realized by some authors \cite{bbp}
\cite{volk},
is responsible 
for the net increase in the probability of interaction per proton 
with respect to the case of a 
straight line propagation. The increase factor can be estimated to be 
$\sim c t_{cl} /R_{cl}\simgt 600$, where $t_{cl} \simlt H_0^{-1}$ 
is the age of the 
cluster and $R_{cl}$ is its size. 
Cosmic rays produced within a cluster during all its 
lifetime can thus produce gamma rays 
through the production and the subsequent decay of neutral pions:
\begin{equation}
p+p\to \pi^0+X~,~~~~~~~~~~~~~~\pi^0\to \gamma+\gamma.
\end{equation} 
Note that 
in the same interactions, charged pions are also produced, which 
determine a neutrino emission through the following channels:
\begin{equation}
p+p\to\pi^{\pm}+X, ~~~~~\pi^{\pm}\to \mu^{\pm} 
\nu_{\mu}(\bar{\nu}_{\mu}), ~~~~~\mu^{\pm}\to e^{\pm} + 
\bar{\nu}_{\mu}(\nu_{\mu})
+ \nu_e (\bar{\nu}_e) ~.
\end{equation} 
We will also discuss the relevance of these last 
processes in Section 7 below.

Using the gamma ray production from clusters of galaxies according
to eq. (1), Houston \ea \cite{hw}
suggested that the total extragalactic 
gamma ray intensity detected above $35$ MeV \cite{ft82},
$I_{\gamma} \approx 5.5 \cdot 10^{-5}~ cm^{-2} s^{-1} sr^{-1}$,
could be ascribed, for a large fraction, to galaxy clusters. 
They predicted a level 
$I_{\gamma} \approx 5 \cdot 10^{-5} cm^{-2} s^{-1} sr^{-1}$ at
energies above $35$ MeV,
assuming an observed local cluster space density, $n_{cl}
\approx 7.3 \cdot 10^{-5} Mpc^{-3}$, integrated out to the Hubble radius,
$R_H=6 \cdot 10^3$ Mpc, 
and neglecting any cosmological effect.
More recently,  Dar \& Shaviv (hereafter DS \cite{ds})
reanalyzed the problem in the light
of the EGRET data \cite{owz94} and calculated the 
contribution to the DGRB from CR interactions in the 
intracluster gas, under the assumption that the energy density of CR's 
in clusters is the same as in our own galaxy (universality). 
With this assumption, Dar \& Shaviv \cite{ds}
predicted a level $I_{\gamma}(> 100~ MeV) \approx 1.2 \cdot
10^{-5}$ photons cm$^{-2}$ s$^{-1}$ sr$^{-1}$, which 
could explain the whole amount of the DGRB 
of extragalactic origin. 
In a following paper, Berezinsky, Blasi \& Ptuskin (hereafter 
BBP \cite{bbp})
relaxed the {\it ad hoc} assumption of universality, and estimated 
the CR 
energy density in clusters due to various possible sources of CR, 
using the 
condition of diffusive confinement of CRs. 
In their approach, BBP \cite{bbp} showed that it is impossible 
to fulfill the universality condition with the usual CR sources in 
clusters, emphasizing 
that the DGRB due to the CR interactions in clusters should be a 
small fraction of the total diffuse flux observed by EGRET. 
This conclusion was reached by the previous authors
under the hypothesis that a large fraction 
of the baryons in the universe is contained inside clusters of galaxies
(BBP considered  that clusters are 
a fair sample of the baryons in the universe 
\cite{wf91}, 
\cite{wetal93}, 
\cite{wf95})
assumed to have a homogeneous inner distribution of gas, $n_e
=const$ (here $n_e$ is the IGM electron number density). 
Because of these assumptions, their results 
depend only on overall cosmological parameters like the
baryon fraction in the universe $\Omega_b$,
and on the cluster size. 

Dar \& Shaviv \cite{ds} also predicted the gamma ray fluxes from a few 
nearby  clusters (Coma, Perseus and Virgo): for these three 
clusters they found $\gamma$-ray fluxes in the range
$F_{\gamma}(>100 ~MeV) \approx 5 \div 20 \cdot
10^{-8} cm^{-2} s^{-1}$. 
In particular, the value which they predicted for A1656 (Coma), 
$F_{\gamma}(>100 ~MeV) \approx 5 \cdot 10^{-8} cm^{-2} s^{-1}$, 
is close to - or slightly higher than - 
the upper limits given by EGRET for this source. 
Similar results were obtained for these clusters 
by Ensslin \ea  \cite{ensslin96} assuming a population of
CRs from radio sources located within galaxy clusters 
in almost equipartition with the IGM thermal energy. 

We stress here that 
in all the previous works a uniform IGM density profile 
was assumed. 
Moreover, the cluster population was not assumed to evolve with cosmic
time, and
the same working hypothesis of no-evolution was assumed for the IGM
content of each cluster.

However, X-ray studies of galaxy clusters, 
have shown that these cosmic structures are indeed well
structured, having a gas density profile 
$n(r) \propto [1+(r/r_c)^2]^{- 3 \beta /2}$, with core radii 
$r_c \approx 0.1 \div 0.3 
\hmpc$ and $\beta \approx 0.6 \div 0.8$ (see \eg \cite{jf92}; 
see also \cite{sa88} and references therein).
Beside this, the IGM is indeed evolving as indicated 
by its sensitive
metal enrichment, $Fe/H \sim 0.2 \div 0.5$ (in solar units, see \eg 
\cite{e90}, \cite{ar}),
shown even for the brighter clusters observable at redshifts $z \sim
0.5$ \cite{lm97}.
Nonetheless, 
there is also an increasing debate on the possible evolution of the 
X-ray
luminosity function observed out to $z \simlt 0.5$ with the EINSTEIN
\cite{gio90} \cite{h92}
and ROSAT satellites 
\cite{ebe97} \cite{nich}
and on the possible evolution of the cluster temperature function
\cite{cmv} \cite{eke98} \cite {vl98}.
If an evolution is present in the cluster population this can be, in
fact, understood as a result of two competing effects:
\newline
{\it i)} a luminosity evolution, where the cluster X-ray luminosity,
$L \propto
n^2 T^{1/2} R^3$ (mainly due to thermal bremsstrahlung), 
changes with redshift due to variations in the gas 
mass density, $n \propto f_g \rho_{cl}$ (where the cluster gas mass is
taken to be a fraction
$f_g \equiv M_{gas}/M$ of the total cluster mass), 
and/or changes in the 
IGM temperature $T$ at fixed mass, 
$M \propto \rho_{cl} R^3$ 
(here $\rho_{cl}$ is the cluster total mass density);
\newline
{\it ii)} a change in redshift of the number density, $N(M,z)$,
(usually referred to as mass function, hereafter MF) of
clusters that are found to be collapsed (or virialized) in the mass range $M,
M+dM$ at redshift $z$.

Detailed studies of cluster evolution in X-rays 
(see \eg \cite{cv}, \cite{ob}, \cite{cmv})
considered in fact that a combination 
of the previous mechanisms is responsible for the actual cluster 
evolution when they fit the available data (see \cite{cv} for a  
detailed discussion).

In this paper we predict the amount of high energy, non-thermal,
gamma-ray emission from galaxy clusters using detailed modelling 
of the realistic cluster structure, as well as viable modelling for
the evolution of the IGM and of the cluster MF. 
Based on these phenomenological  
cluster models, we predict the amount of DGRB 
that can be produced in the viable cosmological models: here we consider
flat and low-density (open or vacuum-dominated) CDM models as well as mixed
Dark Matter models with a fraction $\Omega_{\nu} \approx 0.3$ of 
the total density of the universe in form of
massive neutrinos.
We use $h=H_0/100$ km s$^{-1}$ Mpc$^{-1}$ 
throughout the paper unless otherwise specified.

The plan of the paper is the following.
In Sect.2 we briefly summarize the cluster formation hystory in hierarchical
scenarios for structure formation.
In Sect.3 we describe a model for the production of diffuse gamma-ray
emission due to the interaction of CR's 
with the target protons present in the extended, diffuse IGM.
We consider in Sect. 4 different CR sources that can be found in
connection with galaxy clusters. 
We discuss in Sect.5 the correlation between the 
extended gamma-ray emission and the
much better known thermal X-ray emission coming from the IGM.
Based on these properties, we construct a list of predicted
$\gamma$-ray fluxes for a compilation of X-ray clusters with 
detailed informations on their IGM structure, 
IGM temperatures  and X-ray fluxes.
In Sect. 6 we present predictions for the amount of DGRB produced by
galaxy clusters in different cosmological scenarios. 
We briefly discuss in Sect.7 the extended neutrino fluxes emerging from
these objects and their contribution to a possibly detectable 
diffuse neutrino background (hereafter DNB).
Finally, in Sect.8 we discuss our results in the light of the  
current limits obtained from EGRET and in the light of the future 
experiments for gamma-ray and neutrino astronomy.

\section{Cluster formation: theory}
Clusters of galaxies originate from small density perturbations 
at early times \cite{smoot}
that grow under the action of their own gravitational instability.
In the viable cosmological scenarios, cluster 
evolution can be followed adopting a
simple spherical collapse picture according to which a homogeneous, spherical
perturbation 
detaches from the Hubble flow at time $t_m$ given by the 
equation, $t_m=[3 \pi/32 G \rho_b(t_m)]^{1/2}$, 
collapses at time
$t_c\simeq 2 t_m$, and virializes at time $t_v\simeq 3 t_m$
[here $\rho_b(t_m)$ is the cosmological background density at the
time $t_m$ of the maximum expansion of the perturbation, 
see \cite{peeb80} for details].
The relative density contrast of such a perturbation at the 
initial redshift $z_i$,
$\delta_{i,v}$, depends on the cosmological model.
Under the assumption of linear
growth, the density contrast at $t_v$ is
$\delta_v=\delta_{i,v}{\cal D}(t_v)/{\cal D}(t_i)$, where
${\cal D}(t_v)$ is the linear
growth factor in the chosen cosmology (see \cite{cmrv} for details).
For $\Omega_0\rightarrow 1$, $\delta_v$ tends to the standard value
$\delta_v \approx 2.2$, independent of $t_v$.
The actual, non-linear density contrast,
$\Delta = {\rho}/ \rho_b$ (where ${\rho}$ is the
perturbation density and $\rho_b$ is the background density at the time
of virialization) 
of a cluster that virializes at redshift $z$
in a $\Omega_0\le 1$ cosmological model writes:
$\Delta(\Omega_0,z)=18\pi^2/[\Omega_0(H_0t)^2(1+z)^3]$
[in flat, vacuum dominated low-density models $\Delta$ has not an analytical expression (see, \eg
\cite{cmrv} and references therein)]. It
tends to the standard value
$\approx 400$, found in a flat ($\Omega_0=1$) universe.

Detailed imaging of galaxy clusters in the X-rays (see \cite{sa88})
revealed that the IGM is concentrated in the cluster central
region (the core) and its
density decreases rapidly at large distances from the core.
Thus, following Colafrancesco \ea \cite{cmrv}, we relax the assumption of
uniformity, $n = const$,  by considering a 3-D gas density profile:
\be
n(r) = n_c \bigg[ 1 + \bigg({r \over r_c} \bigg)^2 \bigg]^{-3\beta/2}
~,
\label{eq:eq_den}
\ee
where $n_c$ is the central electron number density and $r_c$ is a core radius.

There are various open issues pertaining to the formation of hot
gaseous cores of galaxy clusters. For our purposes here we can adopt the
following simplified approach (see also \cite{cmrv}): 
shortly after a cluster forms and
virializes, a gaseous core forms (probably as a result of tidal galactic
interactions and other gas stripping processes) with the hot gas in
local hydrostatic equilibrium in the potential wells of the cluster.
Analytical studies of the cluster self-similar collapse \cite{bert}
and numerical
simulations \cite{nfw95} \cite{eke97}
indicate that the gas density profile
scales following  the total matter density of the cluster. 
Thus, the central gas density, $n_c$, 
is related to the total density at the cluster centre, $\rho_c$, 
through
$n_c=f_g (\rho_c/m_p) 2/(1+X)$, where $m_p$ is the proton mass and $X=0.69$
is the cosmic Hydrogen mass fraction. 
The mass within the outer radius, taken as $R=qr_c$,  is
$
M(q,\beta) = 3 M_c \omega(q,\beta),
$
where $\omega(q,\beta)= \int_0^q dx x^2 (1+x^2)^{-3 \beta/2}$.
Here $M_c=(4\pi/3)r_c^3\rho_c$.
Because of the
assumed profile, the ratio between the central and mean mass density
of the cluster is $\rho_c/(\rho_b \Delta)=q^3/3\omega(q,\beta)$.
Assuming that the cluster collapse is self-similar, we can infer the mass and 
redshift dependence of the core radius $r_c$ from the equivalence:
$r_c= R_v/q= {1 \over q}(3 M/4 \pi \rho_b \Delta)^{1/3} (1+z)^{-1}$. 
This gives:
\be
r_c(\Omega_0,M,z) = {1.29 \hmpc \over q} \biggl[ {M\over 10^{15} \msunh}
\cdot {\Delta(1,0)\over \Omega_0 \Delta(\Omega_0,z)}\biggr]^{1/3}
{1\over 1+z} ~.  
\label{eq:r_c}
\ee
Hereafter we fix
$q=10$ to recover $r_c$ values consistent with 
the observations \cite{hm85} \cite{jf92}.

Under the standard assumption
of the IGM in hydrostatic equilibrium with the potential well of a
spherically-symmetric virialized cluster, the 
relation between IGM temperature and cluster 
mass is easily obtained by applying the virial
theorem: $T=-\mu m_p U/(3k M)$, where $\mu=0.62$ is the mean molecular 
weight (corresponding to a hydrogen mass fraction of $0.69$), 
$k$ is the Boltzmann constant and $U$ is the cluster potential energy. 
If the cluster is
assumed to be uniform, $U=-(3/5)GM^2/R_v$ and $T=T^{(u)}=(1/5) 
(\mu m_p/
k)GM/R_v$, where $R_v= [3M/(4\pi \rho_b \Delta)]^{1/3}/(1+z)$ is the
cluster virial radius.
In the  case of the density profile in eq. (\ref{eq:eq_den}), 
$U=-(GM^2/r_c) \psi/\omega^2$ and $T= 5 q \psi
T^{(u)}/(3\omega^2)$, where $\psi=\int_0^q y
dy/(1+y^2)^{3 \beta/2} \int_0^yt^2dt/(1+t^2)^{3 \beta/2}$.
Thus the cluster IGM temperature is
\begin{equation}
T = 5.8 ~keV~ (1+z)\bigg({M \over 10^{15}M_{\odot} h^{-1}}\bigg)^{2/3} 
\cdot \bigg[ { \Omega_0 \Delta(\Omega_0,z) \over \Delta(1,0)}
\bigg]^{1/3} ~,
\end{equation}
where the normalization constant, $T_{0}=5.8$ keV, 
is obtained for $q=10$ and $\beta = 2/3$.
For $\beta=1$ we get $T_{0}\approx 8.2$ keV, where  $T_{0}$ is the 
temperature of a  
$ 10^{15}M_{\odot} h^{-1}$ cluster at $z=0$ in a $\Omega_0=1$ universe.

Although it is expected that the IGM mass fraction, $f_g$, 
depends on $z$ and
$M$, little is currently known on the exact form of these dependences.
We adopt the simple parametrization (described in detail in
Colafrancesco \& Vittorio \cite{cv})
which is suggested by models of the IGM evolution driven by entropy
variation in the cluster cores \cite{eh91} \cite{me94} \cite{b97} and/or shock
compression and heating \cite{ccm} \cite{tm97} \cite{c97}:
\begin{equation}
f_g = f_{g,o} \biggl({M\over 10^{15} \msunh }\biggr)^\eta
(1+z)^{-s} \, .
\end{equation}
The normalization to
$f_{g,o} \simeq 0.1 $, is based on a local, rich cluster sample.
Values of $\eta \sim 0.2 \div 0.9$ and  $s\sim 0.5 \div 2$ are 
consistent with the available data.

The previous information about $n(r)$ [see eq.(3)], $T$
[see eq.(5)] and 
the cluster extension, $R=qr_c$, allow to predict
both the gamma-ray and the X-ray 
emissivities which will be discussed in the following sections.

\section{Diffuse High Energy Emission from Clusters of Galaxies}
As recently discussed by BBP \cite{bbp} and Volk et al. \cite{volk}, 
most 
of the cosmic rays produced in clusters of galaxies remain 
confined within the cluster potential wells  and 
produce high energy gamma rays and neutrinos by interactions with the 
intracluster baryonic gas (the IGM).
The confinement of these CR's in the intracluster space is
a crucial mechanism  for maximizing the efficiency of the
$\gamma$ (or neutrino) emission and is
strictly related to the value and to the 
configuration of the magnetic field in clusters.
This determines the 
diffusion coefficient, given by:
\begin{equation}
D(p)=\frac{1}{3} c r_L(p) \frac{B^2}{\int_{1/r_L(p)}^{\infty} P(k) dk}.
\label{eq:diff}
\end{equation}
\noindent
Here $P(k)$ is the power spectrum of the fluctuations in the magnetic field 
and $r_L(p)=pc/(eB)$ is the Larmor radius of a particle with electric charge 
$e$ and momentum $p$.
In this paper we assume that the fluctuations of the magnetic field have 
a Kolmogorov spectrum:
\begin{equation}
P(k)=P(k_0) \left( \frac{k}{k_0}\right)^{-5/3},
\label{eq:spectrum_fl}
\end{equation}
\noindent
where $k_0=1/d_0$, and $d_0$ is the smallest spatial scale at which 
the magnetic field achieves homogeneity. 
The spectrum in eq.(\ref{eq:spectrum_fl}) is normalized as:
\begin{equation}
\int_{k_0}^{+\infty} P(k) dk \approx B^2,
\end{equation}
\noindent
which implies,
\begin{equation}
P(k_0)=\frac {2}{3}\frac{1}{k_0}B^2=\frac {2}{3} d_0 B^2.
\label{eq:norm}
\end{equation}
Finally, the diffusion coefficient for relativistic particles is 
obtained from
eqs. (\ref{eq:diff}) and (\ref{eq:norm}):
\begin{equation}
D_{CR}(E)=\frac{1}{3} c d_0^{2/3} (eB)^{-1/3} E^{1/3} ~,
\label{eq:diff_1}
\end{equation}
where $E=pc$ is the energy of the relativistic particles.

Measurements of the overall magnetic field in single galaxy clusters yield values 
in the range $\sim 1 \div 10~\mu G$ (\cite{v86}
\cite{v87}
\cite{kk90}
\cite{ketal90} 
\cite{detal87}), 
while a statistical analysis of a sample of clusters 
(see \eg \cite{ld82}) 
yielded typical, average values 
$B\sim 1\mu G$ on 
homogeneity scales $d_0 \simgt 20~kpc$.
This size is compatible with the model of Jaffe \cite{j80}
 for the origin of the 
magnetic field in clusters: according to  this model, the turbulence of the 
IGM produced by the large scale motions of the galaxies is 
responsible for the value and the homogeneity scale of the magnetic field.

Thus, from eq. (\ref{eq:diff_1}) we obtain:
\begin{equation}
D_{CR}(E) \approx 2.3\times 
10^{26}~E(eV)^{1/3}~B_{\mu G}^{-1/3}~cm^2/s  ~.
\end{equation}
The diffusion time of particles with energy $E$ at distance $r$ 
from the CR source reads: 
\begin{equation}
\tau=\frac{r^2}{6D_{CR}(E)} 
	\approx 6.9\times 10^{21} r_{Mpc}^2 E(eV)^{-1/3} s, 
\end{equation}
where the factor $6$ in the previous expression for $\tau$ 
comes from the solution of the
diffusion equation in spherical geometry (here we assumed $B_{\mu G}=1$).
This implies a confinement of cosmic rays, on typical scale of $R_{cl}\approx 
2~Mpc$, if
\begin{equation}
E \simlt 4.2\times 10^{14} h^{-3}~ eV ~.
\label{eq:e_max}
\end{equation}
These cosmic rays have diffusion times larger than the age of universe, for 
which we adopted the value $t_0=2.06\times 10^{17}h^{-1}~s$.

More difficult is to extrapolate, from the average confinement ability of the
clusters, the confinement efficiency of the cluster core: 
the difficulty comes
from the small size of this region ($r_c \sim 0.2\div 0.5~Mpc$) where a
detailed knowledge of the structure
of the magnetic field is needed (the typical separation
between galaxies in clusters cores can be even smaller that the 
size of the largest galaxies often observed in the cores).
 
The confinement of cosmic rays inside clusters and the relevance of this 
process for the production of high energy radiation was already pointed out by 
BBP.  In that work, however, 
the true density  profile of clusters was not 
considered, and only an average IGM density entered the calculation. 
Indeed, for realistic IGM density profiles, like that in eq.
(\ref{eq:eq_den}), a large fraction of cosmic rays could be 
confined inside the dense cluster cores, thus determining 
a sensitive increase in the interaction rates.

Following the approach of BBP, and assuming  a power-law 
spectrum of CR produced by a source located 
at the center of the cluster, 
the number of i-secondaries (i$=\gamma,~ \nu$) produced per 
unit time and unit volume, at energy 
$E$ and at distance $r$ from the CR source in a cluster, is:
\be
q_i(E,r) = Y_i \sigma_{pp} n^2(r) c \bigg[ {n_p(E,r) \over n(r) } 
\bigg] ~,
\ee
where $\sigma_{pp} \approx 3.2 \times 10^{-26} cm^{-2}$, $Y_i$ are 
the yields for the i-secondary production \cite{b90},  
$n(r)$ is the IGM density 
profile given by eq.(\ref{eq:eq_den}) and 
the produced CR proton density, $n_p(E,r)$, is determined from the diffusion
equation as:
\be
n_p(E,r) = { 1 \over 4 \pi r} \bigg[ {Q_p(E) \over D_{CR}(E) } \bigg]
~.
\label{eq_crden}
\ee
In this last equation, $Q_p(E)$ is the emitted spectrum of CR from a source assumed to
be located in the cluster core.
Thus, the total number of i-secondaries produced per unit time at energy $E$ reads:
\be
Q_i(E) = 4 \pi \int_0^{R} dr r^2 q_i(E,r)=
Y_i \sigma_{pp} n_0 c  \bigg[ {Q_p(E) \over D_{CR}(E) } \bigg] r_c^2 
\zeta(q,\beta,E,z;M) ~,
\label{eq:eq_totsec}
\ee
where 
\be
\zeta(q,\beta,E,z;M) = \int_0^{R_{diff}/r_c} dx ~x (1+x^2)^{-3 \beta /2} 
~.
\label{eq:eq_zeta}
\ee
We note that the maximum length over which CR diffuse is 
$R_{diff} \equiv \ell_D =
\sqrt{6 D_{CR}(E) t(z,\Omega_0)}$, at each epoch $t(z,\Omega_0)$.
This yields
\be
\ell_D \approx 0.5 \hmpc \bigg({ t_0 \over 2 \cdot 10^{10} yr} \bigg)^{1/2}
\cdot \bigg( { D_{CR} \over 10^{29} cm^2} \bigg)^{1/2} ~.
\ee 
For a uniform cluster we reproduce the results of BBP.
In the present approach, instead, the inclusion of a realistic density profile 
for the IGM produces a change from a pure power--law CR spectra,
$Q_i(E) \propto E^{-\gamma_g}$ [if $Q_p(E) \propto E^{-\gamma_g}$ 
is assumed], 
due to the presence of an energy dependent term in the function 
$\zeta \propto [1+(\ell_D / r_c)^2]^{-3 \beta/2 +1}$ in
eq.(18).
However, at the energies for which the CR 
confinement is mostly effective,
$\ell_D^2 \ll r^2_c$,  
expanding eq.(\ref{eq:eq_totsec}) in power series, yields:
\be
Q_i(E) \approx 3 Y_i \sigma_{pp} n_c c t(z)  Q_p(E) ~.
\label{eq:eq_totsecapprox}
\ee
In this limit, the total number of i-secondaries produced per unit
time is independent on the cluster size and depends only on the
IGM density, $n_c$ at the cluster centre.

Note that the term $Q_p(E)$ can take into account the possible 
evolution of the luminosity of the CR source as well as the 
possible evolution of the number of CR sources that are present in the central regions of the clusters.
Another important source of evolution in the $\gamma$ (or neutrino) diffuse emission 
from galaxy clusters comes from the possible evolution of  
the IGM content (see Sect. 2).
We will discuss these points in the following.

\section{Sources of Cosmic Rays in Clusters of Galaxies}

In the previous sections we discussed the confinement of cosmic rays 
in galaxy clusters, irrespective of the way they are produced.
In this section we shall deserve more attention to the possible 
CR sources and we will estimate their contribution in terms of 
CR luminosity.

\subsection{Normal galaxies}
The most natural candidates as sources of CR in clusters are normal 
galaxies. If we assume that our own galaxy is a typical one, the 
CR luminosity 
can be estimated to be $L_{CR}\simeq 3\times 10^{40}~erg/s$
\cite{b90}. 
The central number of galaxies
\footnote{
The central number density of galaxies $N_0$ is related to the
total number of galaxies in a cluster by the expression
$N_g = 4 \pi N_0 r_c^3 \omega(q,\beta)$.
}
$N_0$ (which is a measure of the cluster richness) 
is found to be correlated with the
cluster temperature and it scales as
$
N_0 \propto T^{0.8}
$
\cite{e90}.
Thus the gamma ray
luminosity associated with normal galaxies in clusters scales as
$L_{CR} \propto N_0 L_p  \propto T^{0.8} L_p$,
and for a Coma-like cluster with $T=8.3 $ keV,  we find
$L_{CR} \approx 2 \cdot 10^{42}$ erg/s. 
In any case, even for the richest and hottest clusters observed 
(A2163 with
$T=13.9$ keV) the CR luminosity associated to normal galaxies is
found to be $L_{CR} \simlt 10^{43}$ erg/s.

\subsection{Active galaxies}
As discussed by BBP, on statistical grounds we can 
expect 
$\sim 1$ {\it active galaxy} per cluster, where the term {\it active galaxy} 
is used here to indicate several classes of luminous objects, 
such as AGN, 
radiogalaxies and cD galaxies. 
The mechanisms of CR production in active 
galaxies have been studied by many authors (see \eg \cite{b96}
and references
therein) and typical luminosities of $L_{CR} \sim 
10^{44}~erg/s$ can be achieved. 

In a recent paper,  Ensslin et al. \cite{ensslin96}
proposed that jets of powerful radiogalaxies in clusters can
establish a 
sort of equipartition between the thermal energy of the IGM and 
the cosmic ray energy density in clusters, 
due to particle acceleration in the jets. 
The previous authors calculated 
the gamma ray flux from single clusters due to the CR interactions 
in the case  mentioned above and
predicted a gamma ray flux of $F_{\gamma}(> 100 MeV) 
\approx 6\times 10^{-8}~cm^{-2}s^{-1}$ for A1656
(Coma), a little bit in excess of the EGRET experimental limit. Similar
predictions are given for other clusters (A426, Ophiucus, 
M87-Virgo) for which
they found 
$F_{\gamma}(> 100 MeV)$ in the range $\approx 3\div 12 \times 10^{-8}~cm^{-2}s^{-1}$.

Given all these estimates, a CR luminosity $L_{CR} \approx 10^{44}$ erg/s can be
expected from active galaxy interaction with the IGM of local clusters.
At higher redshifts, the AGN luminosity (or the CR density) is expected to
increase roughly as $\sim (1+z)^2$ while the IGM density 
(the target proton
density) is expected to decrease roughly as $\sim (1+z)^{-1.2}$. The two
evolutionary effects tend to balance out and the net result is a moderate
increase, if any, of the effect due to distant, AGN populated clusters.

\subsection{Cluster accretion shocks}

Another important acceleration mechanism for CR's is provided by
accretion shocks formed in and around
clusters of galaxies as a result of their collapse and virialization 
\cite{bert} \cite{ev90} \cite{kj96} \cite{cola97b}.
These shocks have been recognized as possible sources of CR acceleration
through the first order Fermi mechanism (see \eg \cite{bbp} 
\cite{kj96}).

We shall estimate here the CR luminosity by adopting two models 
for the accretion of matter around galaxy clusters. 
In the first model we refer to the analytical calculations of 
Bertschinger \cite{bert}, in 
which a self-similar analytical solution of the hydrodynamic equations is 
found. The solution, concerning the velocity and density of the accreting 
gas, 
is written in terms of the dimensionless variable, $\lambda=r/R_{ta}$, 
where $r$ 
is the distance from the center of the cluster (assumed to have spherical 
symmetry) and $R_{ta}$ is the turn-around radius, given by
\begin{equation}
R_{ta}=\left( {8GMt_0\over \pi^2}\right)^{1/3},
\end{equation}
\noindent
(here $M$ is the cluster gravitational mass). 
In these calculations
\cite{bert} the stationary shock is located at 
$\lambda_{sh} \approx 0.347$. At the shock  
location, the gas 
velocity is found to be $v(\lambda_{sh}) \approx 1.5R_{ta}/t_0$, 
while the density of the gas 
is $\rho_{gas}(\lambda_{sh})\approx 2.2\times 10^{-29}~g/cm^3$. 
This last number has been  obtained 
assuming $\rho_{gas}(r)=0.1\rho(r)$, where $\rho$ is the 
total mass density of the cluster
(dark matter and baryonic gas have been assumed to have the same 
radial distribution).
Thus, the total energy per unit time available at the shock 
can be written as:
\begin{equation}
L={1\over 2} \rho_{gas}(R_{sh}) v(R_{sh})^2 4 \pi R_{sh}^2 v(R_{sh}) \approx
2.4\times 10^{45} ~{\rm erg/s}~ \left( {R_{ta} \over 5 Mpc}\right)^5 
\end{equation}
(we use here $h=0.75$).
The typical efficiency of CR acceleration at the shock is of 
order $\sim 0.1$ \cite{b90}, so that 
$L_{CR}\approx (2\div 3)\times 10^{44}~erg/s$ is obtained, 
which is comparable with the CR luminosities derived in the previous
Section 4.2
for the case of active galaxies within galaxy clusters.

An alternative way to estimate $L_{CR}$ for the accretion shock 
case is to write it through the
matter accretion rate, $\dot{M}$, at the shock.
Numerical simulations of the large scale structure formation 
(\cite{norman} \cite{roett})
give larger values for the shock size, of order $\sim 5~h^{-1} Mpc$. 
If we again assume that a 
fraction $\sim 0.1$ of the gravitational energy at the shock is converted 
in cosmic rays accelerated at the shock, we can write:
\begin{equation}
L_{CR}\approx 0.1 {G M \dot{M}\over R_{sh}}\approx 
(1.3-2.5)\times 10^{44}
~erg/s.
\end{equation}
\noindent
In this numerical estimate,
we assumed that the cluster accretion rate at the shock 
is given by the time averaged value
\begin{equation}
\dot{M}\approx <\dot{M}>={M_{gas}\over t_0}\simeq 5\times 10^{29}~g/s ,
\end{equation}
\noindent 
where $M_{gas}$ is the mass of the cluster in the form of IGM, 
assumed here to be 
$\sim 10\%$ of the total mass.
Note that this last estimate is probably a lower value for the actual
IGM mass in rich clusters (see \eg \cite{wetal93} \cite{david97}).

To summarize, the CR luminosity obtained in the two
accretion shock models yield comparable 
results for the local cluster CR luminosity, $L_{CR}\approx$ 
a few $10^{44}~erg/s$.
At higher redshifts, the CR luminosity from cluster shocks should tend to
decrease, if $f_g$ does.

On the basis of the previous considerations we are justified in assuming an 
average CR luminosity  of $L_{CR}=10^{44}~erg/s$ 
for a typical, rich cluster with 
$M=M_{15}$ (here $M_{15} = 10^{15} \msunh$)
at $z = 0$, 
either due to active 
galaxies and/or to accretion shocks around clusters.

\section{The correlation between X-ray and Gamma-ray
emission}
From eq.(17) we can now calculate gamma-ray fluxes and luminosities for
clusters with mass $M$ at redshift $z$.
In Fig.1 we plot the cluster $\gamma$-ray luminosity against 
the cluster temperature $T$ expected in different, viable  
cosmological scenarios (see Fig.1 caption for details). 
In our specific predictions we find that the cluster $\gamma$-ray 
luminosity $L_{\gamma}$ [or equivalently $Q_i$ as written in eq.(17)]
scales with the cluster properties (at fixed redshift $z$) as:
$$
L_{\gamma} \propto \bigg[ { Q_p(E) \over D_{CR}(E) } \bigg] \cdot n_0 
r_c^2 \zeta ~,
$$
that yields the scaling:
\be
L_{\gamma} \propto \bigg[ { Q_p(E) \over D_{CR}(E) } \bigg] T^{1+\eta}
\zeta ~.
\ee
We assumed here a virial equilibrium for the IGM which yields, $T
\propto GM/R$.
In our specific  model for the IGM
we also consider a distribution of values for
$f_g = M_{gas}/M \propto M^{\eta}$ (with $\eta \sim 0.2 \div 0.9$)
from groups to galaxy clusters in order to recover phenomenologically
the general trend indicated by the most recent data \cite{david97}
\cite{alvio97} \cite{cola97}.
The function $\zeta$ decreases with increasing cluster mass
(temperature) and hence the actual scaling of $L_{\gamma}$ with the
cluster temperature reads $L_{\gamma} \sim T^{0.5}$ that is weaker with
respect to the scaling of the previous equation (see also Fig.1).
The shape of the $L_{\gamma}(T)$ curves  also depends slightly on the
considered cosmological model, with low $\Omega_0$ values providing
higher $L_{\gamma}$ at fixed $T$ (see Fig.1).
This mild cosmological dependence is caused by the sensitivity of the
cluster parameters (mainly the core radius $r_c$) to the underlying
cosmology: in fact, $r_c$ increases with decreasing $\Omega_0$
[see eq.(4)].

In Fig.2 we plot the expected $\gamma$-ray fluxes at $E > 100$ MeV
for clusters with X-ray flux $F_X(2-10 keV)$ in the energy band $2-10$
keV.
We show this plot for clusters at $z=0.023$, the redshift of Coma.
For this cluster 
we have an upper limit on $F_{\gamma}(>100 ~MeV)$ measured by
EGRET (as indicated in Fig.2). 
We plot the fluxes expected in the cosmological models considered in
Fig.1.
The $\gamma$-ray luminosities shown in Fig.1
correspond to fluxes $F_{\gamma}(>100 ~MeV) \simlt 
10^{-9}$ photons s$^{-1}$ cm$^{-2}$ for clusters at $z \sim 0.1$, and 
$F_{\gamma}(>100 ~MeV)$ in the range  
$ 7 \cdot 10^{-9} \div 1.5 \cdot 10^{-8}$ photons s$^{-1}$ cm$^{-2}$ 
for Coma-like 
clusters (at $z \sim 0.023$) in different cosmological models, 
as shown in Fig.2. 
These fluxes are a factor $\sim 2 \div 5$ smaller than the EGRET 
upper limit   
for Coma; such a limit was saturated  by the predictions of 
Dar and Shaviv \cite{ds} and Ensslin \ea \cite{ensslin96}.
Smaller fluxes are obtained for flat ($\Omega_0=1$) cosmologies.
The differences in $L_{\gamma}$ expected in different cosmological
models (see Fig.1) are amplified when one compares the relative fluxes,
$F_{\gamma}= L_{\gamma}/4 \pi d^2_L$, by the $\Omega_0$ dependence of
the luminosity distance $d_L$.
At redshifts $z \simgt 0.3$ the predicted $F_{\gamma}(>100 ~MeV)$ for clusters
fall below values $\sim 10^{-10}$ photons s$^{-1}$ cm$^{-2}$, 
and become hardly observable from any 
present and/or planned gamma-ray mission for the next coming years
(INTEGRAL, GLAST, AMS).

From the previous results, we found a correlation between the X-ray luminosity of galaxy
clusters  and their gamma (or neutrino) luminosity due to CR
interaction in the intracluster gas, as shown in Fig.2. 
Such a correlation is indeed expected because 
the protons in the IGM are both the target for the CR
interactions and the responsible for the thermal bremsstrahlung 
radiation which
provides the bulk of the cluster X-ray luminosity.
In fact, the total X-ray luminosity from a cluster with an IGM temperature $T$ 
can be  written as: 
\be
L_X = A(T) (kT)^{1/2} \int_0^{R} dr 4\pi r^2 n^2(r)~.
\ee
Using eq. (\ref{eq:eq_den}) for the IGM density profile, eq.(26)
writes as:
\be
L_X = 4 \pi n_c^2 r_c^3 A(T) (kT)^{1/2} {\cal I} (q,\beta) ~,
\ee
where
\be 
{\cal I} (q,\beta)=\int_0^q dx x^2 (1+x^2)^{-3\beta}~.
\label{eq:eq_lx}
\ee
The function $A(T)$ in eqs. (26-27) contains 
the contribution of the
Gaunt factors to the frequency integrated X-ray spectrum, 
and writes as:
\be
A(T) = \frac{2^5 \pi e^6 }{3 h m_e c^3} \left( \frac{2 \pi}{3 m_e }
\right)^{1/2} \int dy ~g_{ff}(T,y) e^{-y} ~,
\ee
where $y=h\nu/kT$, $e$ and $m_e$ are the electon charge and mass,
respectively, and $g_{ff}$ is the Gaunt factor.

Using eqs. (8) and (16) we can derive the
expression for the ratio of the the gamma (or neutrino) flux coming 
from a single galaxy cluster and its corresponding X-ray flux, $F_X$:
\be
\frac{F_{\gamma,\nu}}{F_X} = 
\frac{Y_{\gamma,\nu} \sigma_{pp} c}{4 \pi A(T) {\cal I} (q,\beta)}
\left( \frac{1}{n_c r_c} \right) \frac{1}{(kT)^{1/2}} 
\left\{ \frac{Q_p(E)}{D_{CR}(E)} \zeta \right\}.
\label{eq:eq_fluxratio}
\ee
Thus, from the previous eq.(\ref{eq:eq_fluxratio}) it is possible to estimate the
cluster $\gamma$ (or neutrino)
flux given a measurement of the cluster X-ray  flux $F_X$, 
the central IGM density
$n_c$, the cluster core radius $r_c$ and the cluster IGM 
temperature $T$:
\be
F_{\gamma, \nu} = F_X \cdot C(E,\beta,z) {1 \over n_c r_c \sqrt{kT}}
\ee
Here the quantity $ C(E,\beta,z)$ reads:
\be
C(E,\beta,z)= 
\frac{Y_{\gamma,\nu} \sigma_{pp} c}{4 \pi A(T) {\cal I} (q,\beta)}
\left\{ \frac{Q_p(E)}{D_{CR}(E)} \zeta \right\}.
\ee
We will use here the previous results and the available 
data for $F_X, n_c, r_c, \beta$ and $T$ \cite{jf92} \cite{david93} 
\cite{ha91}
to derive a list of estimated cluster fluxes in the 
high energy $\gamma$ and neutrino energy regions (see Table 1).
The clusters in the sample listed in Tab.1 
are those which have well measured values
of $\beta, n_c, r_c, T$ and $F_X$ and are selected from the
homogeneous sample of Jones and Forman \cite{jf84} which is limited to
redshifts $z \simlt 0.1$.
In Fig.3 we plot these data in the $F_{\gamma}(>100 MeV)-T$ plane,
together with the predicted curves, $F_{\gamma}(>100 MeV)(T)$, 
expected 
in a CDM+$\Lambda$ model with $\Omega_0=0.4$ and $h=0.6$ at redshifts
$z=0.023$ (solid line) and at $z = 0.072$ (dashed line).
The theoretical predictions at the minimum and maximun redshifts
of the clusters in the Jones \& Forman sample
encompass the data points.
The $\gamma$-ray fluxes predicted from eq.(30-31) 
for our sample of
clusters at $z \simlt 0.1$ span over the range
$F_{\gamma}(>100 MeV) \approx 5 \cdot 10^{-10} \div  8.5 \cdot 10^{-9}$
photons s$^{-1}$ cm$^{-2}$  and are distributed (with some intrinsic
dispersion) along the theoretical $F_{\gamma}(>100 MeV)-T$ curves
predicted at each redshift (see Fig.3) according to eq.(31).
Note that the region at $T\simlt 2$ keV is populated mainly by galaxy
groups for which the determinations of the IGM temperatures are more
difficult and uncertain due to their small amount of diffuse IGM
(see \cite{c98a} \cite{alvio} for a discussion).
In the following, we will consider mainly galaxy clusters with $T
\simgt 2$ keV.

In Fig.4 we show how the clusters listed in Tab.1 
distribute according to their predicted $\gamma$-ray flux.
The distribution of the cluster $\gamma$-ray fluxes peaks at
$F_{\gamma}(>100 MeV) \sim 8 \cdot 10^{-10} \div  3 \cdot 10^{-9}$
photons s$^{-1}$ cm$^{-2}$.
Our sample of clusters was extracted from an X-ray flux limited sample
\cite{jf84} and thus it can be considered as representative of the
cluster population even in the $\gamma$-rays because of the
existing correlation $F_{\gamma}-F_X$ of eq.(32).
From Fig.4 we note that the optimal observational strategy to
detect large samples of galaxy clusters in the 
$\gamma$-rays is to achieve
sensitivities $\simgt 5 \cdot 10^{-10}$ photons s$^{-1}$ cm$^{-2}$:
such a sensitivity level should be at hand with the next generation
$\gamma$-ray telescopes (INTEGRAL, GLAST).

\section{The contribution of galaxy clusters to the DGRB}

The diffuse flux of $\gamma$ (or neutrino) emission at energy $E_i$ that is 
received from a cluster at redshift $z$ is given by
\be
F_{\gamma, \nu} (E_i)= { Q_i[E_i(1+z),z] \over 4 \pi {d}_L^2 } 
\label{eq_gflux} ~,
\ee
where the luminosity distance
\be
{d}_L(z, \Omega_0) = {c\over H_0} (1+z) \int_0^z dz'~ 
\bigg[(1+z')^3 \Omega_o +1 -\Omega_o \bigg]^{-1/2}~,
\label{eq:eq_dl}
\ee
allows for the flux--luminosity conversion:
\be
L=4\pi F {d}_L^2(z, \Omega_0) .
\ee
The total flux of $\gamma$-rays (or neutrino) due to the superposition of a 
population of evolving clusters is then:
\be
I_{\gamma, \nu} (E') = \int_0^{z_{max}} {dV(z,\Omega_0) \over dz} dz
\int_{M_{min}}^{\infty} dM N(M,z) { Q_i[E_i(1+z),z] \over 4 \pi
{d}_L^2 } ~,
\label{eq:eq_grbflux}
\ee
where $dV(z,\Omega_0)/dz$ is the cosmological 
volume element per unit steradian.

The space density of clusters at any mass $M$ and redshift $z$ is 
given by the mass function, $N(M,z)$, (hereafter MF)
usually derived by the  Press \& Schechter \cite{ps74}  
theory as:
\be
N(M,z) = \sqrt{2\over\pi} {\rho_b\over M^2} {\delta_v \over
\sigma}{dln
\sigma 
\over dln M}\exp[-\delta_v^2/2\sigma^2] ~.
\ee
We remind the reader that
$\rho_b$ is the comoving background density of the universe, 
$M$ is the total cluster mass and 
$\delta_v$ is the linear density contrast of a perturbation that
virializes at redshift $z$ (see Sect.2). 
The  variance $\sigma$  of
the (linear) density  fluctuation field at the scale  $R=(3 M / 4\pi
\rho_b)^{1/3}$ and redshift $z$  is given by the standard relation:
\be
\sigma^2(R,z)= {{\cal D}^2(\Omega_0,z) \over 2 \pi^2} 
\int k^2 dk P(k) \bigg({3 j_1(kR) \over kR}\bigg)^2
\ee
\cite{peeb80},
where ${\cal D}$ is the growth factor of linear density fluctuations in a
given cosmology and $j_1(x)$ is a spherical Bessel function. 
If we normalize the fluctuation spectrum, 
$P(k)$, by requiring
$\sigma(8 \hmpc,z=0) = b^{-1}$, then the MF depends uniquely on the product $b
\delta_v$, for a given cosmological model. 

Assuming a power-law power spectrum for the initial fluctuation field, 
${\cal P}(k) = A k^n$, the MF attains the following analytical expression:
\be
N(M,z)= \sqrt{{2 \over \pi}}
{\rho_b\over M_0^2}{n+3\over 6} {\delta_v b \over {\cal D}(z)}
\bigg({M\over M_0} \bigg)^{\alpha -2}
\exp\bigg[-{1 \over 2} {\delta_v^2 b^2 \over {\cal D}^2(z)} \bigg({M\over M_0}
\bigg)^{2\alpha} \bigg] \, ,
\label{eq:mass_distr}
\end{equation}
where $\alpha=(n+3)/6$ and $M_0$ is the mass contained in a $8 \hmpc$
radius sphere at $z=0$.
The MF we will use here is normalized to the observed abundance 
of clusters at $z \approx 0$
by fitting the parameter $\delta_v b$ (see \cite{cv} \cite{cmv}).
This fitting procedure is done by using the available data on the temperature
function \cite{ha91} and/or those on the X-ray luminosity function 
\cite{ko84} \cite{ebe97}.
The different data sets yield approximately the same best fit values 
for $\delta_v b$ in the cosmological models we considered.
Specifically, CMV \cite{cmv} 
found $\delta_v b= 2.5$ for a standard CDM model and 
$\delta_v b= 1.6$ for a low density, vacuum dominated CDM model with
$\Omega_0=0.4$ and $\Omega_{\Lambda}=0.6$, which is the best fitting CDM model.
For this class of CDM models, CMV found  
$\delta_v b \approx 2.5 \Omega_0^{0.49}$, or equivalently 
$\sigma_8 \approx 0.67 \Omega_0^{-0.49}$ (if one assumes $\delta_v=1.68$).

In the case of power-law power spectra and
for $\Omega_0=1$, eqs. (\ref{eq:eq_dl}), (\ref{eq:eq_grbflux}) and 
(\ref{eq:mass_distr}) give the following analytical result for the
cluster contribution to the DGRB:
$$
I_{\gamma}=\sqrt{\frac{2}{\pi}} \frac{\rho_b}{4\pi} M_0^{-1/3}
\left( \frac{n+3}{6}\right) (\delta_v b)  Y_{\gamma}
\sigma_{pp} n_c \frac{c^2}{H_0} \frac{Q_p(E)}{D_{CR}(E)} 
\chi^2
$$
\be
\times \int_0^{z_{max}} dz
(1+z)^{-(\gamma_g+ \epsilon +2-5/2)}
\zeta
\int_{y_{min}}^{\infty} dy y^{\alpha-2+2/3}
exp\left\{ -\frac{1}{2} 
\left(\frac{\delta_v b}{{\cal D}(z)}\right)^2 y^{2\alpha} \right\},
\ee
\noindent
where $y=M/M_0$, $\chi=r_c(1+z)/M^{1/3}$ has been obtained 
from eq. (\ref{eq:r_c}), 
$\zeta$ is defined in eq. (\ref{eq:eq_zeta}) and 
$\epsilon$ is the power index of the energy 
dependence of $D_{CR}(E)$: 
$D_{CR}(E)\propto E^{\epsilon}$, where we found 
$\epsilon=1/3$ (see eq.11).

A numerical calculation of the integrals in eq.(40) 
yields the immediate result that the main
contribution to the $\gamma$ (or neutrino) diffuse flux comes from 
low-$z$ clusters,  and actually
from sources at $z\simlt 0.3$. 
This result is confirmed by our 
numerical calculations 
of the redshift distribution, $N(z)$, 
of $\gamma$-ray emitting clusters found 
in the different cosmological models we considered in this paper.
In Fig.5 we show the redshift distributions, $N(z)$, as a function of
$z$, of $\gamma$-ray clusters
expected in a flat ($\Omega_0=1, h=0.5$) CDM model in the case of an
IGM evolving as $f_g \propto M^{0.2}(1+z)^{-1.8}$ (solid
histogram) and in the case of absence of IGM evolution with redshift
(dotted histogram).
We considered clusters with fluxes $F_{\gamma}(>100 MeV) > 10^{-10}
~photons s^{-1} cm^{-2}$ with $\gamma_g=2.1$.
The presence of an IGM evolution modifies the cluster distribution
mainly at $z \simgt 0.2$.
Models with no IGM evolution (\ie with $s=0$) predict $\sim 10$ times
more clusters with $F_{\gamma}(>100 MeV) \simgt 10^{-10} photons
s^{-1} cm^{-2}$ at $z \simgt 0.4$ with respect to the case of a strong
evolving IGM model with $s=1.8$.
However, such a difference tends to reduce to $\sim 30 \%$ at $z \simlt
0.2$ and becomes negligible at $z \simlt 0.1$.
As the bulk ($\simgt 90 \%$) of the bright $\gamma$-ray clusters is
located at $z \simlt 0.15$, the IGM evolution should play a little
role on the cluster contribution to the DGRB. 

The value of $z_{max}$ in eq.(40) is defined by two criteria: 
i) the largest cutoff in the redshift distribution of
the cluster population; 
ii) the maximum energy of the i-secondaries. In fact, for
$E \simlt E_{max}$, the energy of production of the i-secondaries is 
$E(1+z) \simlt E_{max}$  (that is the maximum energy of the 
original CRs).
If $E > E_{max}$, we must realize that the spectrum of the 
i-secondaries is changed.

Note that in these reasonings, the diffusion coefficient $D_{CR}(E)$ depends 
on the typical homogeneity scale $d_0$ of the magnetic field in clusters
(see Section 3).
There are reasonable arguments to expect that this scale could change
with redshift.
However, we assume here that the only dependence with $z$ of the diffusion 
coefficient $D_{CR}$ is due to its dependence on the cosmic time 
(see eqs.11-13).

In Fig.6 we show the cluster contribution to the DGRB, 
$I_{\gamma}$, as a
function of the $\gamma$-ray energy $E$,
in the various cosmological models we consider in this paper (see
Fig.6 caption for details).
The expected contribution of galaxy clusters to the high 
energy DGRB [evaluated according to eqs.(36-40)] 
is found to be $\simlt 5 \%$ of the EGRET value
\cite{owz94}
for all the considered cosmological models (see Fig.6).
This result depends on different issues:
\newline
{\it i)} we choose $L_{CR}=10^{44}$ erg/s for a $M_{15}$ cluster
in each cosmological model.
The value of $I_{\gamma}$ depends linearly on the choice of  
$L_{CR}$ so that any variation of this parameter reflects in a
analogous variation of the predicted contribution to the DGRB;
\newline
{\it ii)} we normalize the local abundance of clusters to that
observed in the X-rays. This estimate is the most precise and robust at the
moment. Such a normalization yields a values of $\delta_v b$ for each
cosmological model, so that the evolution of galaxy clusters is
completely determined;
\newline
{\it iii)} the distribution of the cluster IGM is self-consistently
derived from the gravitational instability picture for the formation
of galaxy clusters. The level of $I_{\gamma}$ depends linearly on the
value of $f_{g0}(M_{15},z=0)$ for which we choose the value $0.1$.
Changing this parameter in the viable range,  $f_{g0} \approx 0.05 -
0.3$ yield similar variations in $I_{\gamma}$;
\newline
{\it iv)}
only low-$z$ clusters effectively contribute
to the integral in eq.(36) as shown by the redshift distribution
of clusters with $F_{\gamma} \geq 10^{-10}~cm^{-2} s^{-1}$ (see Fig.5).
Nonetheless, there is a mild dependence of the results from the
considered cosmological model.
The MDM models show the smallest value of the DGRB contributed
by clusters due to the very recent formation epochs predicted for clusters in
this model (see Fig.6).
CDM models (flat and low density) show approximately the same level of
DGRB: this is due to the normalization of their spectral amplitude to the
observed abundance of nearby clusters observed in the X-rays
\cite{cmv}.

In Fig.7 we show the quantity $I_{\gamma}$ as a function of the
$\gamma$-ray energy $E$, expected in a CDM$+\Lambda$ model
($\Omega_0=0.4, h=0.6$) for the cases of the maximum rate of IGM
evolution allowed by the present data, $f_g \propto M^{0.2}
(1+z)^{-2.2}$ (solid line) and for the case $f_g=$ const (dashed
line).
We note that the amount of DGRB contributed by galaxy clusters
in different cosmological models
depends slightly of the IGM evolution (see Fig.7)
as the clusters effectively
contributing to the DGRB are those with $z \simlt 0.2$.
At these redshifts,  the maximum change of $f_g$ (at fixed mass) 
is of the order of $\sim 20 \%$
and this corresponds to a variation of $I_{\gamma}$ of the same order
(see eqs.18, 36 and 40).

Variations in the parameters of the cluster structure yield variations
$\simlt 5\%$ on the final result. Changing $\beta$ by $20 \%$ yields
variations in $\zeta$ of $\sim 8 \%$ that give similar changes on 
$I_{\gamma}$.
Variations in the primordial index $n$ of the perturbation spectra
yield variations of order $\sim (n+3)/6$ on $I_{\gamma}$.

Thus, while the level of the DGRB predicted for a specific choice of
the cluster evolution models are at the level of $0.5\div 0.8 \%$ of the
EGRET diffuse flux, more conservative estimates of $I_{\gamma}$
require to consider the theoretical uncertainties in the modelling of
cluster structure and evolution. 

Theoretical uncertainties in the description of the relevant
quantities of cluster structure yield variations of the results 
contained within $\simlt$ a few  \%. Let us be more specific.

Our results are based on the 
theoretical description of cluster evolution through the spherical
top-hat model and the PS mass function.
Despite the inherent uncertainties in the detailed models for cluster
collapse, the final results depend on the combination $b \delta_v$
that is fixed by the fitting procedure to the local cluster abundance
with uncertainties $\sim 5 \div 8 \%$ \cite{cmv}.
Moreover, our results on $I_{\gamma, \nu}$ are weakly sensitive to the
lower cutoff, $M_{min}$, in the MF (see eqs 35-39). 
This reduces the
effects of possible changes of the  slope of the MF in the efforts to
go beyond the simple, first-order theory for the origin of the
universal MF (see \cite{cola97} for a discussion).

Uncertainties in the description of the IGM evolution
(both its normalization, $f_{g0} \sim 0.1\div 0.3$ at the scale of rich
clusters, and its evolution with time parametrized in terms of the
parameters $\eta \sim 0 \div 0.9$ and $s\sim 0 \div 1.5$) are
discussed to point out the role of different, possible gas-phase
phenomena in cluster evolution.
Here we take into account the minimal and maximal levels of IGM
evolution that are still consistent - given the observational
uncertainties - with the current data (\cite{david93} \cite{gio90}
\cite{h92} \cite{nich}).

So, in Fig.8 we show the possible range of $I_{\gamma}$ 
produced by a population of galaxy clusters considering 
reasonable uncertainties in the theoretical description of the clusters
structure and of their distribution in mass 
for the considered cosmological models.
According to our previous estimation of the uncertainties, it is
reasonable to expect that galaxy clusters could provide up to a few
$\%$ of the EGRET \cite{owz94} DGRB. Higher values of $I_{\gamma}$ would
require a choice $L_{CR} \gg 10^{44}$ erg/s, or a large 
contribution from galaxy groups: both these possibilities seem to be
quite unrealistic.

\section{High energy neutrinos from clusters of galaxies}

Gamma rays from galaxy clusters are always accompanied by neutrino 
production
due to the decay of charged pions, according with eq.(2).
In Fig.9 we show the range of cluster contribution to the 
diffuse neutrino background
(DNB), $I_{\nu}$,  as a
function of the neutrino energy $E$, expected in the viable models for
structure formation.
We considered in the predictions of Fig.9 the level of theoretical
uncertainties in the description of the cluster structure and evolution
(see Sect.6 for a discussion).
We considered the whole contributions due to the fluxes of $\nu_{\mu}$
and $\nu_e$, and we compare them to the level of the atmospheric
neutrinos as given by Gondolo \ea \cite{gondolo}.
The diffuse $\nu$ flux produced by clusters is at a 
level of $F_{\nu} \simgt 10^{-2}~neutrinos~ km^{-2}$ $yr^{-1} sr^{-1} 
GeV^{-1}$   
at $E \simgt 3 \cdot 10^5$ GeV, where the 
spectral shape of the cluster fluxes ($\propto E^{-2.1 \div 2.3}$) 
emerges from
the $E$ distribution of the atmospheric $\nu$'s (see Fig.9).
However,  such an estimate has to be considered quite optimistic
because it refers to the specific choice of parameters that maximize
$I_{\gamma}$.
In such a maximal model for $I_{\gamma}$, 
%
%
%
at $E=10^3$ GeV one
could observe a sensitive diffuse flux from galaxy clusters at a level 
$I_{\gamma} \sim 30 \div 2\cdot 10^3~ neutrinos ~km^{-2} yr^{-1} sr^{-1}
GeV^{-1}$. 
This diffuse emission could be one of the most intense neutrino
backgrounds of cosmological origin at these energies, comparable only
to that produced by unresolved AGN cores \cite{gais97}.
Single clusters could be  detected with the next
generation, large neutrino telescopes
as the signals from rich, nearby Coma-like clusters are of the
order $F_{\nu} \approx 10^3 \div 10^4 ~neutrinos
~km^{-2} yr^{-1}$ (at $z \simlt
0.05$) at $E \sim 10^3$ GeV.
Much distant clusters are more difficult to be detected as their
fluxes scale as $\propto d_L^{-2}(z, \Omega_0)$
[see eqs.(34-35) and Fig.2].

\section{Discussion and conclusions}

In this paper we presented a detailed study of the diffuse 
emission of
$\gamma$-rays and neutrinos from clusters of galaxies.
Using realistic modelling of the cluster structure, of their formation
history and of their evolution with cosmic time, we found that
galaxy clusters can provide $\simlt 1\%$ of the DGRB 
measured by EGRET (in the first release by OWZ \cite{owz94}).

Our estimate of $I_{\gamma}$ 
is quite independent on the geometry of the universe, 
on the assumed cosmological model and on the amount of 
IGM evolution, because most of
their contribution to the DGRB comes from nearby, $z \simlt 0.2$,
clusters.
In fact, at these redshifts the effects of curvature do not take 
place strongly in changing the perturbation growth factor, 
${\cal D}(z,\Omega_0)$, (normalized at the present epoch),
the difference in cluster evolution are small when the
different models are normalized to the local abundance of clusters
observed in X-rays and the available amount of IGM evolution - 
even if considered to be quite strong, $f_g \propto (1+z)^{-1 \div -2}$ - 
can provide only small variations to the cluster $\gamma$-ray
luminosities, as $L_{\gamma} \propto f_g$ (see eq.18).
On account of all these aspects, we consider that our results for the
contribution of galaxy clusters to the DGRB are quite robust.

Our approach  differs substantially from the previous ones in several
(among others) aspects:
\par\noindent
{\it i)} we considered - differently from all the previous approaches - 
a self-consistent approach to the formation of
clusters following the spherical collapse model \cite{peeb80}
complemented with a realistic IGM density profile, 
consistent with the most
recent determinations from X-ray observations.
This fact has important effects on the CR confinement within cluster
cores and hence on the relative $\gamma$-ray and neutrino emission rates;
\par\noindent
{\it ii)} we considered (as BBP did) here an energy dependent
diffusion coefficient which results in a very general
picture of the CR confinement within cluster cores;
\par\noindent
{\it iii)} we also considered here - at variance with the previous
approaches - the effects of a possible evolution in the cluster IGM 
content. This is consistent with the present indications of a variation in
the IGM content from groups to rich clusters in the local frame and
with the X-ray, shock (or entropy) induced, luminosity evolution observed from numerical
simulations \cite{tm97}
and predicted in analytical models (both shock
and entropy models) for the 
evolution of X-ray clusters \cite{cola97} \cite{b97};
\par\noindent
{\it iv)} we use the PS cluster MF that was found to be 
consistent with N-body
simulations over a large dynamical range and up to $z \simgt 2 $ 
\cite{eke96}.
We normalized it to the local abundance of clusters
detected in X-rays.
In the previous approaches average values for the overall
cluster abundance, $n_{cl} \approx 4\div 7 \cdot 10^{-5}~ Mpc^{-3}$, 
were
used \cite{hw} \cite{bbp} without considering the effect of an
evolving cluster mass function;
\par\noindent
{\it v)} using a self-consistent modelling of the IGM we found
an
analytical correlation between $\gamma$-ray and $X$-ray emission
for clusters.
The predicted ratio $F_{\gamma} / F_X \propto f_g^{-1}
r_c^{-1} T^{-1/2}$  [see eq.(32)] provides a 
behaviour of the $F_{\gamma}-F_X$ relation different from that
obtained by Ensslin \ea 
\cite{ensslin96},
$F_{\gamma} / F_X \propto T^{1/2}$, because we did not assume any
(partial) equipartition between IGM thermal energy and relativistic
jet particles.
Our correlation results only from the basic electromagnetic and
hadronic emission mechanisms in which the IGM protons are the targets
for both the X-ray thermal bremsstrahlung emission and for the
$p p$ collisions responsible for $\gamma$-rays.
\par\noindent
{\it vi)} using such a correlation we derived a sample of nearby
clusters  with predicted $\gamma$-ray fluxes observable with the next
generation $\gamma$-ray telescopes.
Incidentally, we found a $\gamma$-ray flux for Coma, 
$F_{\gamma}(>100 MeV)\approx 8.5 \cdot 10^{-9}~ photons$ $s^{-1}
~cm^{-2}$ which is consistent with the EGRET upper limits for this
cluster (previous specific predictions \cite{ds}
\cite{ensslin96} seem to exceed the EGRET upper limit).
 
Our numerical results for $I_{\gamma,\nu}$ are in reasonable agreement
with those obtained by BBP, even though based on a quite different
description of the cluster structure and evolution.
This agreement is due to the fact that BBP considered a constant
comoving density of clusters, 
$n_{cl} \sim 5 \cdot 10^{-5}$ Mpc$^{-3}$,
assumed to be  a fair sample of the baryons in the universe, 
and containing a fraction 
$\Omega_b\approx 0.5 \Omega_{BBN}$ (where $\Omega_{BBN}$ is the value
of the baryon density predicted by Big Bang Nucleosynthesis).
Under these assumptions, BBP obtained a value for $I_{\gamma}$ higher by
a factor $\sim 3\div 4$ with respect to our result, based on values $f_g \sim 0.1$
(see Sect.2).
Our refined calculations show why their assumption of considering
$n_{cl} \sim const$ was reasonable:  the clusters
effectively contributing
to the DGRB are located at $z \simlt 0.2$ (see Fig.5), 
where the effects of
evolution do not have room to take definitely place (see Fig.7).

Because of the inherent uncertainties in the predictions of quantities
whose calculation involve to set the values of parameters which are
not known precisely, we also estimated the range spanned by
$I_{\gamma, \nu}$ for the
combination of parameters allowed by the present observational ranges.
In fact, the description of the cluster structure and evolution that we used
in our analytic approach consider only ensemble averaged quantities.
But we observe a whole distribution of
the real cluster properties
with respect to the average cluster moulding.
Some amount of variance is needed to be considered in 
cluster modelling in order to ensure the predictive power of 
 the viable models for structure formation. 
To explore the role of the uncertainties in the relevant
quantities  we considered several sources of uncertainties (see
Sect.7).

From an inspection of Figs. 9 and 10 we note that the effects of  the possible
theoretical uncertainties in the description of the cluster
and IGM evolution
could change the predicted contributions for
the DGRB and for the DNB by a factor $\simlt 3$, setting the maximal
level of $I_{\gamma}$ to a few $\%$ of the EGRET value.

The DGRB seems to be mostly produced by AGNs (FSRQ  and/or BLLacs)
and/or blazars (we take here an  estimate of $\sim 60 \div 65 \%$ 
\cite{erl96} of the EGRET diffuse flux \cite{owz94}).
Diffuse $\gamma$-ray emission could be observed also  
from normal galaxies yielding a contribution $\sim 5 \%$
\cite{erl96}.
When added to the $\sim 10 \div 15 \%$ of the DGRB contributed
by their high-$E$ photons
interacting with 
other existing backgrounds (\eg IR, CMB
\cite{erl96}) one gets  only $\sim 15 \div 20 \%$ of the
DGRB left for truly diffuse or extended sources.
Of this amount, 
a fraction of the diffuse $\gamma$-ray flux $\sim 3 \div 5 \%$ is predicted 
\cite{ww}
to originate 
from decaying topological defects (see \cite{sigl}) and interactions of
UHE particles with the CMB.
Note, however, 
that the amount and the spectral distribution of this possible 
diffuse background depend
sensitively on the amplitude of the primordial magnetic field 
on scales larger than supercluster sizes.
So, according to the previous estimates, the presence of all these  
sources of diffuse $\gamma$-ray
emission (even though partially model dependent)
determines an upper limit to the contribution of extended 
extragalactic sources to the DGRB, that is $\sim 10 \div 22 \%$
of the OWZ EGRET level \cite{owz94}.
This sets  rather weak
constraints on the level of CR production in clusters and hence on the
presence and activity of AGNs in clusters or on the formation and 
efficiency of accretion shocks around clusters.
However, 
if we consider the revised level of the DGRB as derived by SWZ
\cite{swz}, then the previous upper limit reduces to $\simlt 2 \%$.
Our predictions of the DGRB contributed by galaxy clusters 
$I_{\gamma,cl} / I_{EGRET} \sim 0.005 \div 0.02$
is
perfectly compatible with the presence of both a population of
evolving FSRQ and AGNs dominating the $\gamma$-ray sky and with the
presence of truly diffuse backgrounds like those previously discussed.
Note, however, that the major source of uncertainty in the level of
the extragalactic DGRB comes from the contribution of the AGNs.
A fluctuation 
analysis of the EGRET data is needed to have 
more definite indications on the
level of the DGRB contributed by discrete sources.
If, on the other hand, CR acceleration will be found to be relevant 
in clusters (yielding $L_{CR}$ substantially  larger than 
$ 10^{44}$ erg/s), then the
predicted level of DGRB produced by galaxy clusters can set 
interesting limits to the space density and evolution of 
$\gamma$-ray AGNs.

The sensitivities and angular resolutions achievable by the next 
generation gamma-ray (INTEGRAL, GLAST, AMS) and
neutrino (see \cite{halz} for a review) detectors will be able to shed a new
light on the high energy phenomena occurring in large scale
structures.

\vskip 1.5truecm
\newline
{\bf Acknowledgements} 
We thank the Referee for useful comments and suggestions which
improved substantially the presentation of the paper.
We also aknowledge interesting and stimulating 
discussions with V.S. Berezinsky during a recent visit of S.C. at the
LNGS.
S.C. aknowledges also interesting discussions with G. Kanbach,
A. Dar and F. Halzen, among others, 
at the 1997 Moriond Meeting {\it High Energy Phenomena in
the Universe}.
The research of P.B. is funded by a INFN PostDoctoral Fellowship at
the University of Chicago.

\newpage


\newpage

\medskip
\begin{center}
{\bf Table 1.} A list of predicted $\gamma$-ray fluxes from clusters
of galaxies.
\end{center}
\begin{center}
\begin{tabular}{|l|l|l|l|l|l|l|l|}
\hline
Cluster & $z$ & $n_c(^a)$ & $T$(keV) & $r_c$(Mpc) & $\beta$ & 
$F_X(^b)$ &  $F_{\gamma}(^c)$ \\ 
\hline
& & & & & & & \\
A1060 & 0.0111 & 4.55 & 3.9 & 0.1   & 0.67 & 0.488  & 0.457 \\
A262  & 0.0168 & 4.45 & 2.4 & 0.095 & 0.55 & 0.224  & 0.149 \\
A426  & 0.0183 & 4.55 & 6.3 & 0.285 & 0.57 & 7.50   & 1.136 \\
A1367 & 0.0213 & 0.95 & 3.7 & 0.43  & 0.53 & 0.335  & 0.163 \\
A400  & 0.0231 & 1.75 & 2.5 & 0.165 & 0.57 & 0.087  & 0.094 \\
A1656 & 0.0232 & 2.89 & 8.3 & 0.42  & 0.75 & 2.51   & 0.85  \\
A2199 & 0.0305 & 8.8  & 4.5 & 0.14  & 0.68 & 0.694  & 0.234 \\
A2063 & 0.0337 & 4.15 & 4.1 & 0.175 & 0.62 & 0.246  & 0.110 \\
A576  & 0.0392 & 4.05 & 4.3 & 0.115 & 0.49 & 0.197  & 0.059 \\
A2657 & 0.0414 & 3.6  & 3.4 & 0.145 & 0.53 & 0.157  & 0.062 \\
A2319 & 0.0529 & 3.1  & 9.9 & 0.41  & 0.68 & 1.21   & 0.267 \\
A85   & 0.0556 & 5.0  & 6.2 & 0.225 & 0.62 & 0.622  & 0.145 \\
A2256 & 0.0601 & 2.45 & 4.3 & 0.45  & 0.73 & 0.505  & 0.242 \\
A1795 & 0.0621 & 5.8  & 5.8 & 0.3   & 0.72 & 0.519  & 0.130 \\
A1775 & 0.0709 & 4.15 & 4.9 & 0.185 & 0.66 & 0.097  & 0.046 \\
A399  & 0.0715 & 3.05 & 5.8 & 0.215 & 0.52 & 0.342  & 0.078 \\
& & & & & & &  \\
\hline
\end{tabular}
\end{center}
\noindent
{\bf Table caption}
\vskip 0.5truecm
($^a$)~~ $n_c$ in units $10^{-3}$ cm$^{-3}$.

($^b$)~~ X-ray fluxes in the $(2-10)$ keV band in units $10^{-10}$ erg 
cm$^{-2}$ s$^{-1}$.

($^c$)~~ Gamma ray fluxes $F_{\gamma}(>100 MeV)$ in units 
$10^{-8}$ photons s$^{-1}$ cm$^{-2}$.


\newpage
\begin{figure}[thb]
 \begin{center}
  \mbox{\epsfig{file=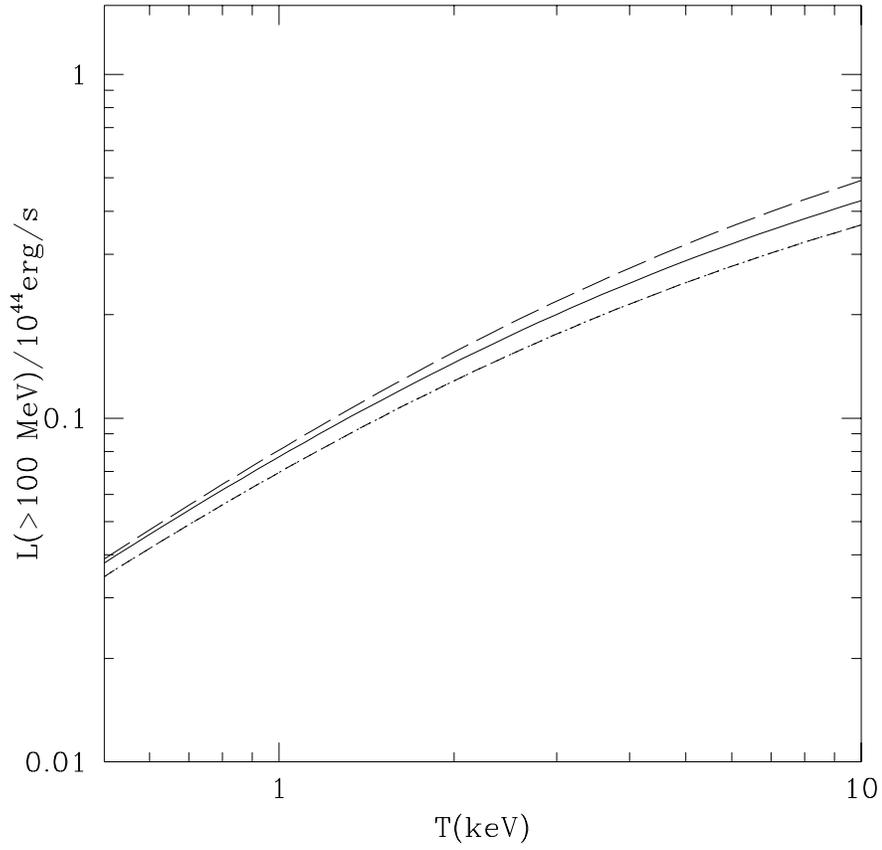,width=12.cm}}
  \caption{\em {Gamma ray luminosities for clusters in
different cosmological scenarios:
CDM ($\Omega_0=1$; dashed curve), CDM$+\Lambda$ 
($\Omega_0=0.4$; continuous curve), open CDM ($\Omega_0=0.3$; 
long-dashes curve)
and mixed DM ($\Omega_0=1$, $\Omega_{\nu}=0.3$ ; dotted curve).
}}
 \end{center}
\end{figure}

\newpage
\begin{figure}[thb]
 \begin{center}
  \mbox{\epsfig{file=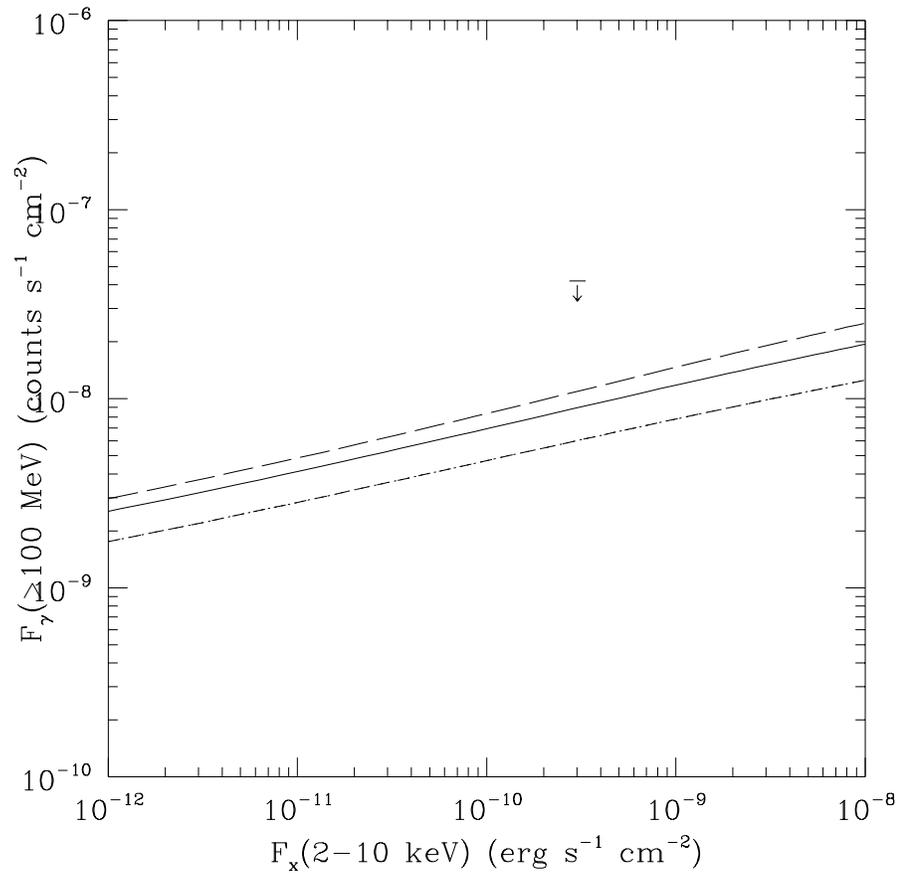,width=12.cm}}
  \caption{\em {
Gamma ray fluxes for clusters at  $z =0.023$, the redshift of the Coma
cluster. 
The arrow indicates the EGRET upper
limit for A1656 (Coma).
Curves labelled as in Fig.1.
}}
 \end{center}
\end{figure}

\newpage
\begin{figure}[thb]
 \begin{center}
  \mbox{\epsfig{file=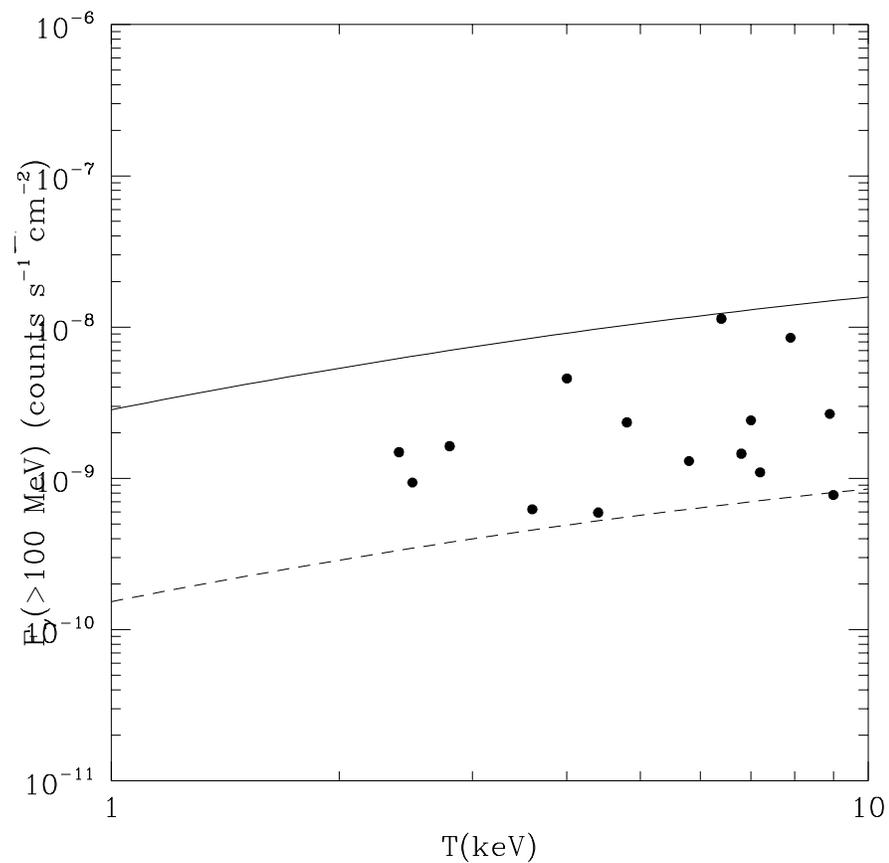,width=12.cm}}
  \caption{\em {Cluster $\gamma$-ray fluxes for the objects listed in
Table 1 plotted against their IGM temperature (filled dots).
The curves represent the $F_{\gamma} -T$ relationship 
in a  CDM$+\Lambda$ 
($\Omega_0=0.4, h=0.6$) cosmology at redshift $z=0.023$ (continuous
curve) and at the maximum redshift of the clusters in our sample (dashed
curve).
}}
 \end{center}
\end{figure}

\newpage
\begin{figure}[thb]
 \begin{center}
  \mbox{\epsfig{file=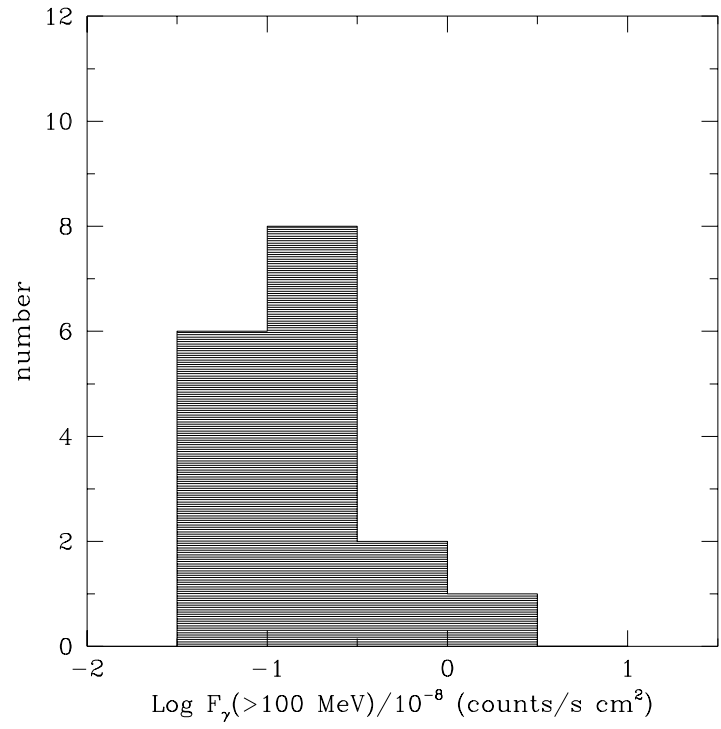,width=10.cm}}
  \caption{\em {The frequency of clusters with a predicted
$\gamma$-ray flux in the sample of Table 1.
}}
 \end{center}
\end{figure}

\newpage
\begin{figure}[thb]
 \begin{center}
  \mbox{\epsfig{file=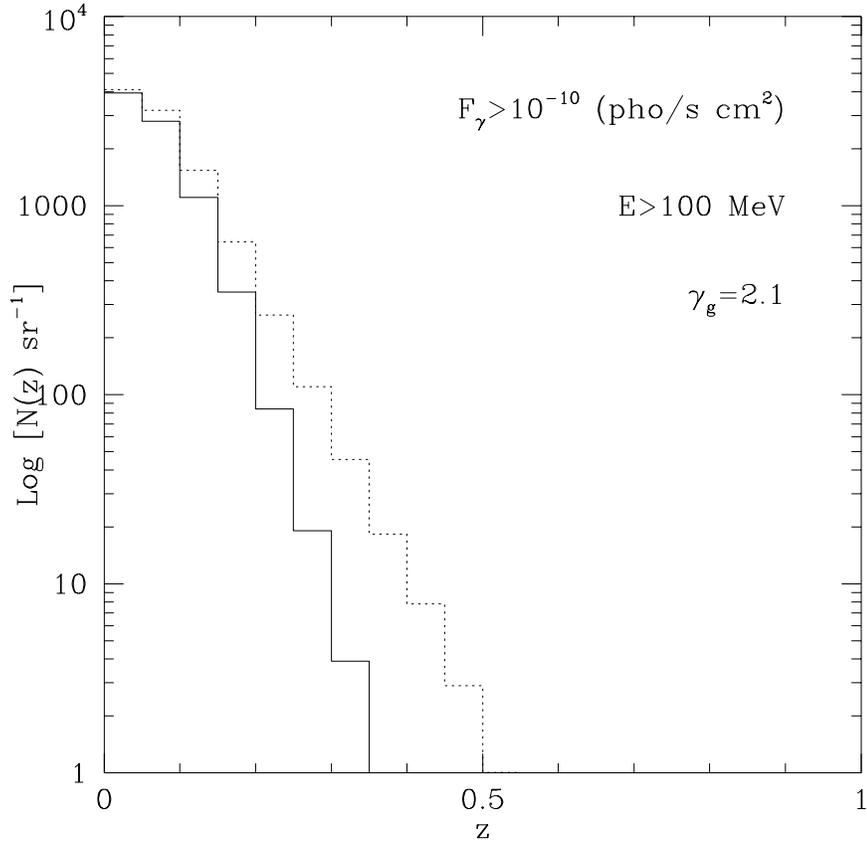,width=12.cm}}
  \caption{\em {
The redshift distribution of clusters with fluxes $F_{\gamma}(> 100
MeV)$ in a CDM model with ($\eta=0.2, s=1.8$: continuous histogram) 
and without ($\eta=0.2, s=0$: dotted histogram) 
IGM evolution (see text for details).
}}
 \end{center}
\end{figure}

\newpage
\begin{figure}[thb]
 \begin{center}
  \mbox{\epsfig{file=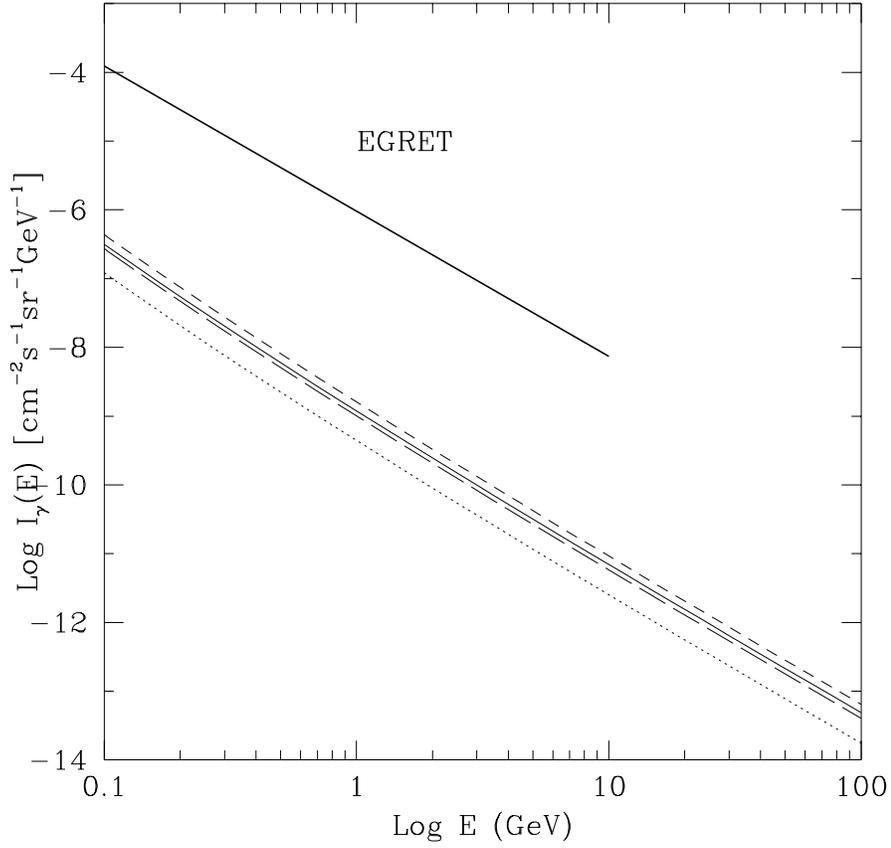,width=12.cm}}
  \caption{\em {
The contribution of galaxy clusters to the DGRB in different
cosmological models:
CDM ($\Omega_0=1$; dashed curve), CDM$+\Lambda$ 
($\Omega_0=0.4$; continuous curve), open CDM ($\Omega_0=0.3$; 
long-dashes curve)
and mixed DM ($\Omega_0=1$, $\Omega_{\nu}=0.3$ ; dotted curve).
}}
 \end{center}
\end{figure}

\newpage
\begin{figure}[thb]
 \begin{center}
  \mbox{\epsfig{file=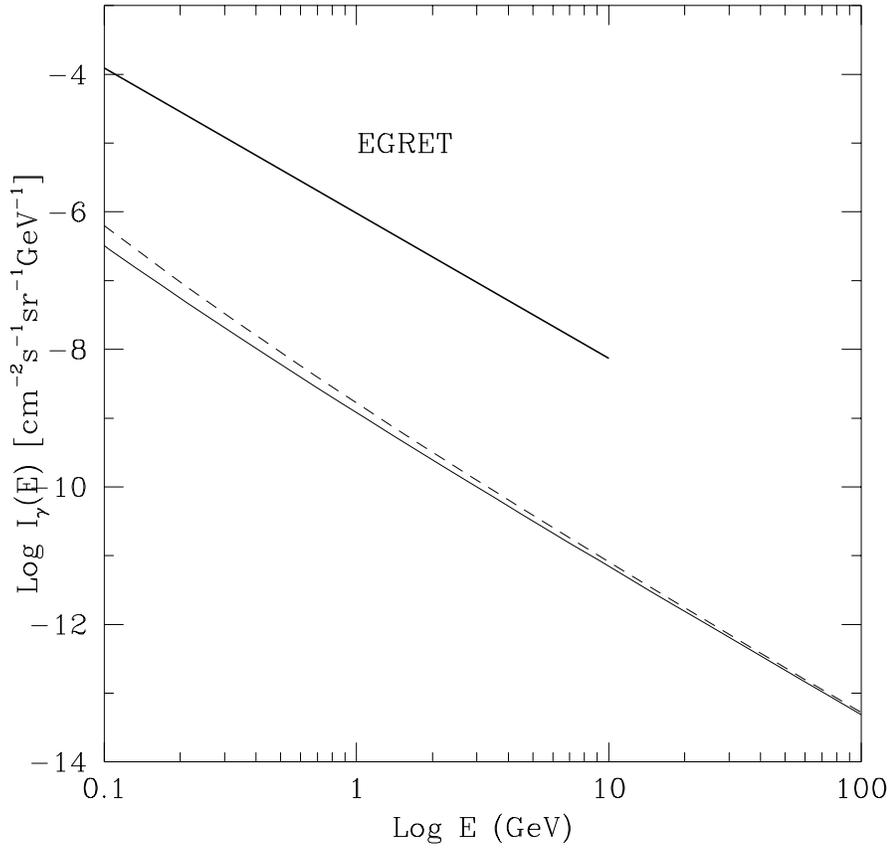,width=12.cm}}
  \caption{\em { 
The predicted level of DGRB from galaxy clusters in a CDM$+\Lambda$
($\Omega_0=0.4, h=0.6$) model 
with ($\eta=0.2, s=2.2$: continuous curve) and without ($\eta=0, s=0$ 
dashed curve) IGM evolution (see text for details).
}}
 \end{center}
\end{figure}

\newpage
\begin{figure}[thb]
 \begin{center}
  \mbox{\epsfig{file=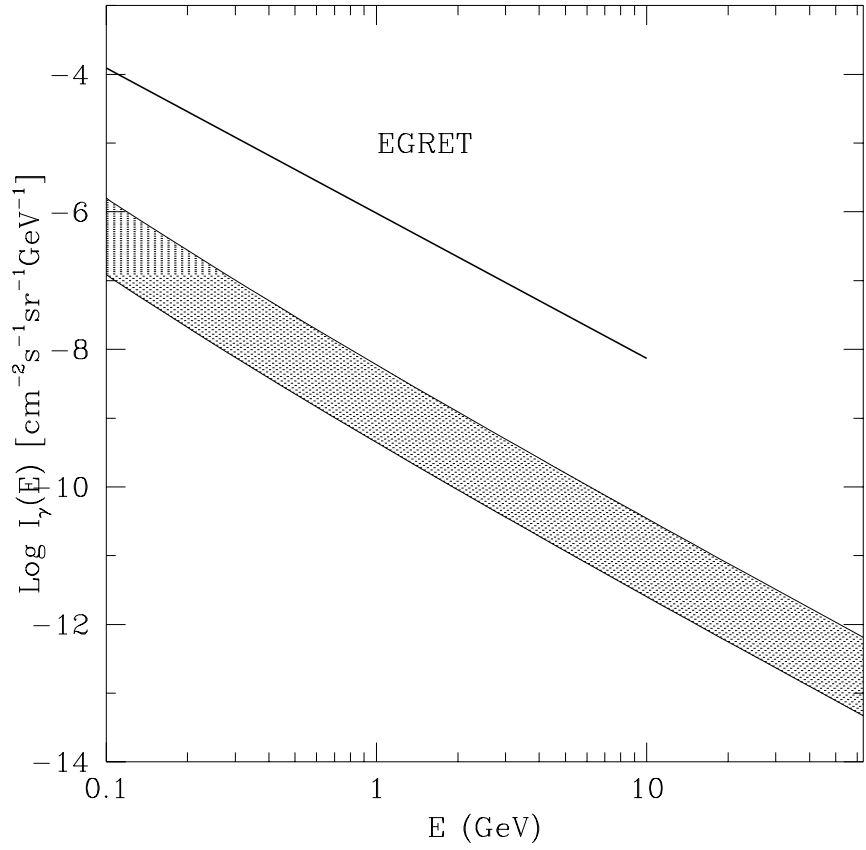,width=12.cm}}
  \caption{\em {
The expected DGRB from clusters,
considering various sources of theoretical uncertainties in the cluster
modelling.
A flat CDM ($\Omega_0=1; h=0.5; n=1$) model is considered here.
}}
 \end{center}
\end{figure}

\newpage
\begin{figure}[thb]
 \begin{center}
  \mbox{\epsfig{file=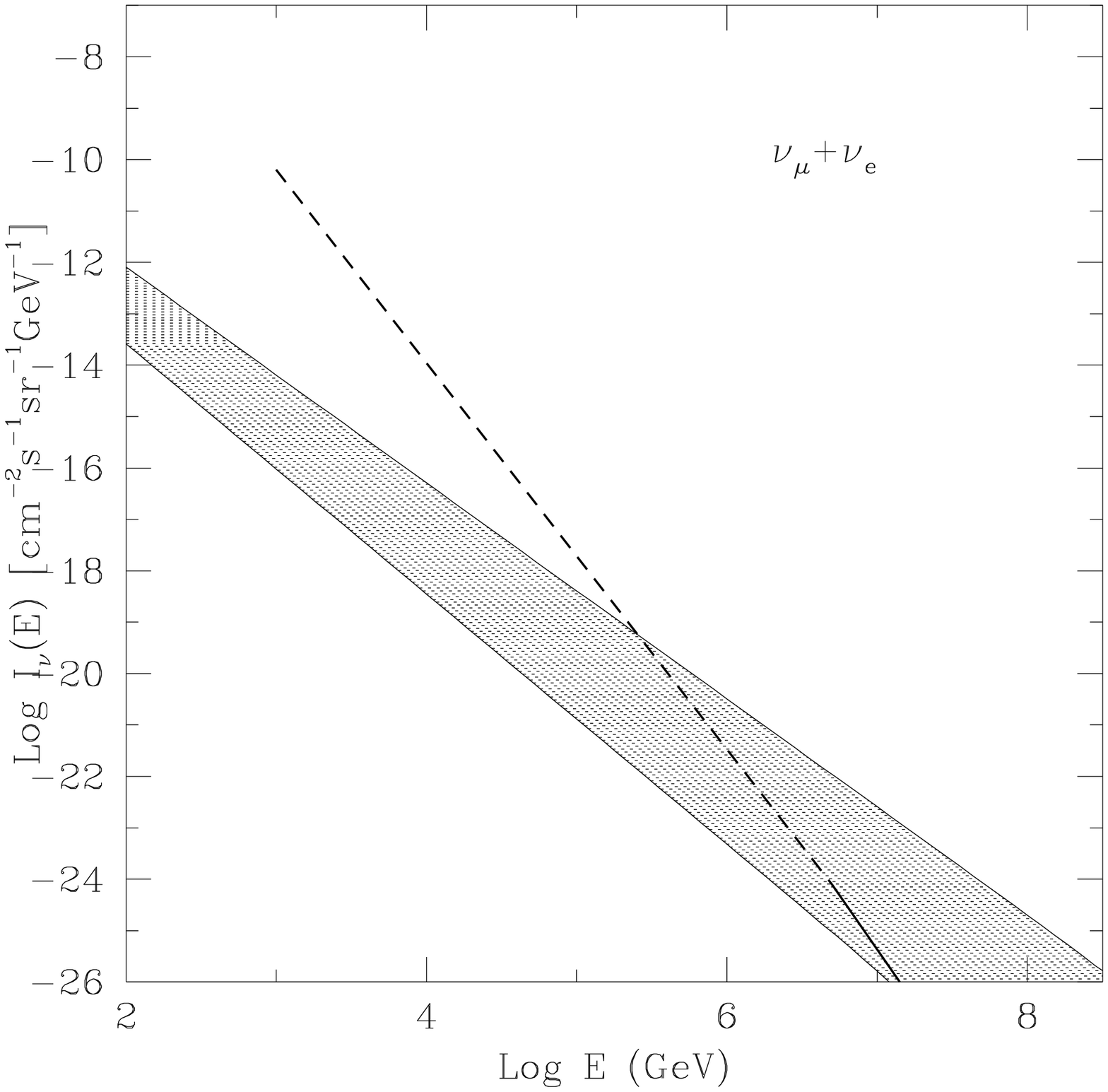,width=12.cm}}
  \caption{\em { 
Same as Fig.8 but for 
the diffuse $\nu$ background.
Heavy dashed line show the diffuse flux from atmospheric $\nu$'s
(Gondolo \ea 1995).
}}
 \end{center}
\end{figure}


\begin{thebibliography}{9}

\bibitem{owz94}
Osborne, J.L., Wolfendale, A.W. and Zhang, L. 1994, J.Phys.G, 20, 1089
(OWZ)

\bibitem{swz} Smialkowski, A., Wolfendale, A.W. and Zhang, L. 1997,
Ap. Phys., 7, 21 (SWZ)

\bibitem{kan} Kanbach, G. 1988, Space Sci. Rev., 49, 69

\bibitem{ca}
Coppi, P.S. and Aharonian F.A. 1996, preprint astro-ph 9610176

\bibitem{fi96}
Fichtel, C. 1996, A\&A Suppl., 120, 23

\bibitem{has}
Hasinger, G.  1996, A\&A Suppl., 120, 607

\bibitem{mat}
Mattox, J.R. \ea 1996, preprint astro-ph9612187

\bibitem{com} 
Comastri, A., DiGirolamo, T. and Setti, G. 1996, A\&A
Suppl., 120, 627.   

\bibitem{pad} 
Padovani, P. \ea 1993, MNRAS, 260, L21

\bibitem{pohl} 
Pohl, M. \ea 1997, preprint astro-ph/9703147 

\bibitem{hw} 
Houston, B.P. \ea  1984, J.Phys.G., 10, L147

\bibitem{ds} 
Dar, A. \& Shaviv, N.J. 1995, Phys, Rev.Lett., 75, 3052 (DS)

\bibitem{bbp} 
Berezinsky, V.S., Blasi, P. \& Ptuskin, V.S. 1996, ApJ, 487, 529 (BBP)

\bibitem{volk}
Volk, H.J., Aharonian, F.A. and Breitschwerdt, D. 1996, Space Sci.
Rev., 75, 279

\bibitem{sa88} Sarazin, C. 1988, {\it X-Ray Emission from Clusters of
Galaxies}, Cambridge Univ. Press

\bibitem{ft82}
Thompson, D.J. and Fichtel, C. 1982, A\&A, 109, 352

\bibitem{wf91}
White, S.D.M. \& Frenk, C.S. 1991, ApJ, 379, 52.

\bibitem{wetal93}
White, S.D.M.  et al. 1993, Nature, 366, 429.

\bibitem{wf95}
White, S.D.M. \& Fabian, A.C. 1995, MNRAS, 273, 72.

\bibitem{ensslin96}
Ensslin, T.A., Biermann, P.L., Kronberg, P.P. \& Wu, X-P., ASTRO-PH/9609190.

\bibitem{jf92} 
Jones, C. \& Forman 1992, in {\it Clusters and Superclusters of
Galaxies}, Ed. A. Fabian \ea (Cambridge: Cambridge Univ. Press), p.49.

\bibitem{e90} 
Edge, A.  \ea 1991, MNRAS, 252, 414

\bibitem{ar}
Arnaud, M. \ea 1992,  A\&A, 254, 49

\bibitem{lm97}
Mushotzky, R. and Lowenstein, M.1997, preprint astro-ph/9702149

\bibitem{gio90}
Gioia, I. \ea 1990, ApJ, 356, L35

\bibitem{h92}
Henry, J.P. \ea 1992, Apj, 386, 408

\bibitem{ebe97} 
Ebeling, H. \ea 1997, preprint  astro-ph/9701179

\bibitem{nich} 
Nichol, R.C. \ea 1996, preprint  astro-ph/9611182 


\bibitem{cmv} 
Colafrancesco, S., Mazzotta, P. and Vittorio, N. 1997, ApJ, 488, 566
(CMV)

\bibitem{eke98} Eke, V., Cole, S., Frenk, C. \& Henry, P. 1998,
preprint astro-ph/9802350

\bibitem{vl98} Viana, P. \& Liddle, A.R. 1998, preprint 
astro-ph/9803244

\bibitem{cv} 
Colafrancesco , S. \& Vittorio, N. 1994, ApJ, 422, 443 (CV)

\bibitem{ob} Oukbir, J. and Blanchard, A. 1997, A\&A, 317, 1

\bibitem{smoot}
Smoot, G. \ea 1992, ApJ, 396, L1

\bibitem{cmrv} 
Colafrancesco, S., Mazzotta, P., Rephaeli, Y., and  
Vittorio, N.  1997, ApJ, 479, 1 (CMRV)

\bibitem{bert} 
Bertschinger, E.1985, ApJS, 58, 39.

\bibitem{nfw95}
Navarro, J.F., Frenk, C. and White, S.D.M 1996, preprint
astro-ph/9611107 

\bibitem{eh91} Evrard, A. \& Henry, P. 1991, ApJ, 383, 95

\bibitem{me94} Metzler, C. \& Evrard, A. 1994, ApJ, 437, 564

\bibitem{b97} Bower, R. 1997, MNRAS, 288, 355

\bibitem{ccm} Cavaliere, A., Colafrancesco, S. and Menci, N. 1993, 
ApJ, 415, 50 

\bibitem{tm97}
Takizawa, M. and Mineshige, S. 1997, preprint astro-ph/9702047

\bibitem{c97} Colafrancesco, S. 1998, in {\it Multifrequency Behaviour
of Cosmic Sources}, Ed. F. Giovannelli, in press

\bibitem{eke97}
Eke, V., Navarro, J.F. and Frenk, C. 1997, preprint astro-ph/9708070

\bibitem{c98a} Colafrancesco, S., Martinelli, A., 
Matteucci, F.,Vittorio, N. and Antonelli, A. 1998, preprint

\bibitem{alvio} Renzini, A. 1997, preprint astro-ph/9706083

\bibitem{hm85} Henriksen, M., and Mushotsky, R. 1985,
ApJ, 292, 441

\bibitem{cola97} 
Colafrancesco, S. 1997, in {\it Frontier Objects in Astrophysics and
Particle Physics}, F. Giovannelli and G. Mannocchi Eds., SIF Conf.
Proc., vol.57, p.71

\bibitem{v86}
Vall\`ee, J.P., MacLeod, J.M. \& Broten, N.W. 1986,
A\&A, 156, 386.

\bibitem{v87}
Vall\`ee, J.P., MacLeod, J.M. \& Broten, N.W. 1987, Astron. Lett. Comm., 25, 

\bibitem{kk90}
Kronberg, P.P. \& Kim, K.-T, 1990, Geophys. Astrophys. Fluid Dyn., 50, 7.

\bibitem{ketal90}
Kim, K.T., Kronberg, P.P., Dewdney, P.E. \& Landecker, T.L. 1990, ApJ, 355, 
29.

\bibitem{detal87}
Dreher, J.W., Carilli, C.L. \& Perley, R.A. 1987, ApJ, 316, 611.

\bibitem{ld82}
Lawler, J.M. \& Dennison, B. 1982, ApJ, 252, 81.

\bibitem{j80} 
Jaffe, W.J. 1980, ApJ, 241, 924.

\bibitem{b90}
Berezinsky, V.S. \ea 1990, {\it Astrophysics of Cosmic Rays},
North-Holland

\bibitem{b96} 
Bierman, P.L. 1996, in {\it High Energy Astrophysics}, Ed.
J.M.Matthews, World Sc., p. 217.

\bibitem{ev90}
Evrard, A. 1990, in {\it  Clusters of Galaxies}, W. Oegerle \ea Eds., 
STScI Symp. Series, 4, p.71

\bibitem{kj96}
Kang, H. \& Jones, W. 1996, preprint astro-ph/9607049.

\bibitem{cola97b}
Colafrancesco, S. 1997, Proc. of the $1^{st}$ Airwatch Symp., 
L. Scarsi \ea Eds., in press

\bibitem{norman}
Norman, M. \ea 1997, preprint 

\bibitem{roett}
Roettiger, K., Burns, J.O. and Loken, C. 1996, ApJ, 473, 651

\bibitem{david97}
David, L.W. \ea 1997, preprint

\bibitem{alvio97}
Renzini, A. 1997, preprint astro-ph/9706083

\bibitem{david93}
David, L.W. \ea 1993, ApJ, 412, 479

\bibitem{jf84}
Jones, C. and Forman, W. 1984, ApJ, 276, 38

\bibitem{ha91} 
Henry, J.P. and Arnaud, K.A., 1991, ApJ, 372, 410

\bibitem{ps74} 
Press, W.H. and Schechter, P. 1974, ApJ, 187, 425

\bibitem{peeb80} 
Peebles, P.J.E. 1980, 
{\it `The Large Scale Structure of the Universe'}, 
(Princeton: Princeton University Press).

\bibitem{ko84} 
Kowalski, M.P. et al., 1984, ApJ S., 56, 403

\bibitem{gais97}
Gaisser, T.K. 1997, preprint astro-ph 9707283

\bibitem{eke96}
Eke, V. \ea 1996, MNRAS, 282, 263

\bibitem{erl96}
Erlykin \ea 1996, A\&A Suppl. 120, 623

\bibitem{ww}
Wdowczyk, J. and Wolfendale, A.W. 1990, ApJ, 349, 35

\bibitem{sigl}
Sigl, G. \ea 1994, Ap. Phys., 2, 401

\bibitem{halz}
Halzen, F. 1997, preprint astro-ph/9703004
 
\bibitem{gondolo} Gondolo, P. \ea 1995, preprint

\end{thebibliography}
\end{document}